\documentclass[manuscript,screen]{acmart}

\usepackage{xcolor}
\usepackage{tcolorbox}
\usepackage{array}
\usepackage{multirow}
\usepackage{tabularx}

\usepackage{xltabular}

\newcolumntype{C}[1]{>{\centering\let\newline\\\arraybackslash\hspace{0pt}}m{#1}}
\usepackage[colorinlistoftodos]{todonotes}
\usepackage[subtle]{savetrees}

\usepackage{tikz}

\newcommand{\circref}[1]{%
  \tikz[baseline=(char.base)]{
    \node[shape=circle, draw, inner sep=1pt, font=\small] (char) {#1};
  }%
}

\AtBeginDocument{%
  }

\setcopyright{acmlicensed}
\copyrightyear{2026}
\acmYear{2026}
\acmDOI{XXXXXXX.XXXXXXX}

\acmConference[Conference acronym 'XX]{Make sure to enter the correct
  conference title from your rights confirmation emai}{June 03--05,
  2018}{Woodstock, NY}
\acmISBN{978-1-4503-XXXX-X/18/06}

\usepackage{tikz}
\usetikzlibrary{arrows.meta,positioning,fit,shadows.blur}
\tikzset{
  box/.style={rounded corners, draw, align=center, inner sep=6pt, font=\small, fill=white},
  titlebox/.style={rounded corners, draw, align=center, inner sep=6pt, font=\normalsize\bfseries, fill=white},
  arrow/.style={-Latex, line width=0.8pt},
  faint/.style={font=\footnotesize, align=center},
  anchcol/.style={fill=blue!8, draw=blue!60},
  expcol/.style={fill=orange!10, draw=orange!70!black},
  devcol/.style={fill=gray!10, draw=gray!60},
  outcol/.style={fill=green!10, draw=green!60!black},
  blur shadow
}

\begin{document}



\title[A Scoping Review \& Theoretical Framework for HCI Design]
{\texorpdfstring{
Grounding Mindfulness in Embodied Tangibles:\\
A Scoping Review \& Theoretical Framework for HCI Design
}{
Grounding Mindfulness in Embodied Tangibles: A Scoping Review \& Theoretical Framework for HCI Design
}}

\author{Tharaka Sachintha Ratnayake}
\orcid{0009-0004-6408-7587}
\affiliation{%
  \institution{The University of Melbourne}
  \country{Australia}
} 
\email{tsratnayakem@student.unimelb.edu.au}

\author{Samangi Wadinambiarachchi}
\orcid{0000-0002-0953-5306} 
\affiliation{%
  \institution{The University of Melbourne}
  \country{Australia}
}
\email{samangi.w@unimelb.edu.au}

\author{Sarah Schömbs}
\orcid{0009-0001-3251-3199}
\affiliation{%
  \institution{The University of Melbourne}
  \country{Australia}
}
\email{sschombs@student.unimelb.edu.au}

\author{Jonathan Eden}
\orcid{0000-0003-0733-265X}
\affiliation{%
  \institution{The University of Melbourne}
  \country{Australia}
}
\email{eden.j@unimelb.edu.au}

\author{Denny Oetomo}
\orcid{0000-0002-2680-6489}
\affiliation{%
  \institution{The University of Melbourne}
  \country{Australia}
}
\email{doetomo@unimelb.edu.au}

\author{Wafa Johal}
\orcid{0000-0001-9118-0454}
\affiliation{%
  \institution{The University of Melbourne}
  \country{Australia}
}
\email{wafa.johal@unimelb.edu.au}

\renewcommand{\shortauthors}{Ratnayake et al.}

\begin{abstract}
Embodied and tangible devices are increasingly used to support mindfulness practices across meditation, yoga, and everyday routines. However, existing HCI research lacks a coherent theoretical foundation for explaining how such systems support distinct mindfulness processes and outcomes. First, we report findings from a scoping review of tangible devices (n=65) for mindfulness in HCI based on the mechanisms of action of mindfulness. The review found that most systems primarily target attentional regulation and body awareness, while emotion regulation and change in perspective on the self remain comparatively underexplored. Also, the evaluation methods used to assess the effectiveness of tangible systems for mindfulness-related outcomes were found to be fragmented and weakly grounded in theory. Building on these findings, a theoretical framework grounded in the Self-Awareness, Self-Regulation, and Self-Transcendence (S-ART) framework is proposed to explain how tangible systems support different forms of mindfulness practice. Within the framework, we present two complementary models: (1) Embodied Sensory Expansion (ESE) for focused-attention meditation (FAM) and (2) Embodied Sensory Anchoring (ESA) for open-monitoring meditation (OMM). Together, we show how this framework can provide a principled basis for explaining designs, generating hypotheses, and evaluating tangible mindfulness technologies in HCI, while highlighting key gaps for future research.
\end{abstract}

\ccsdesc[500]{Hardware~Tactile and hand-based interfaces}
\ccsdesc[500]{Hardware~Haptic devices}
\ccsdesc[500]{Human-centered computing~Haptic devices}
\ccsdesc[300]{Human-centered computing~User interface toolkits}
\keywords{Do, Not, Us, This, Code, Put, the, Correct, Terms, for,
  Your, Paper}


\maketitle

\section{Introduction}
An estimated $\sim 4\%$ of the global population experiences anxiety~\cite{javaidEpidemiologyAnxietyDisorders2023}, and 4.4\% suffer from depression~\cite{friedrichDepressionLeadingCause2017}, making these mental health disorders among the most prevalent worldwide~\cite{jalaliGlobalPrevalenceDepression2024}. In response, many individuals turn to mindfulness-based interventions (MBIs)~\cite{vandamMindHypeCritical2018, khouryMindfulnessbasedStressReduction2015}. Mindfulness is a psychological process of ``bringing attention to the present moment without judgment''~\cite{kabat_zinnMindfulnessbasedStressReduction2003}. It is developed from Buddhist tradition, often cultivated through meditation to foster insight (Vipassanā) and compassion (mettā/karuṇā)~\cite{bodhiWhatDoesMindfulness2013}. 
The connection between mindfulness and meditation is particularly important, as research has demonstrated that repeated momentary experiences of mindful awareness (\textit{state mindfulness}), obtained through meditation, can lead to a more enduring tendency to be mindful (\textit{trait mindfulness}) in everyday life~\cite{miaoRelationshipEmotionalIntelligence2018}.

Recognizing this potential, the Human-Computer Interaction (HCI) community has, over the past three decades, increasingly developed \emph{mindfulness-enhancing technological systems}. In ~\citet{terzimehicReviewAmpAnalysis2019}'s review they identified three types of technological systems: mobile applications~\cite{vaccaDesigningInteractiveLoving2016, zassmanMindfulScrollInfinite2024}; extended Reality (XR) experiences~\cite{gromalaVirtualMeditativeWalk2015, kosunenRelaWorldNeuroadaptiveImmersive2016}; and tangible devices~\cite{ezerSomaestheticMeditationWearable2024a, daudenroquetInteroceptiveInteractionEmbodied2021a} to be major implementations of the design space. Here, tangible devices refer to purpose-built tools specifically designed to support mindfulness through integrated sensing and actuation mechanisms. Owing to their design, these devices are more likely to employ haptic, tactile and embodied modalities compared to mobile or XR systems. While each method has both advantages and disadvantages, with no single approach being clearly superior, we focus in this work on tangible devices because their inherently body-based interactions align closely with mindfulness's strong emphasis on somatic experience ~\cite{shustermanBodyConsciousnessPhilosophy2008}.

Tangible meditation devices have increasingly been commercialized with examples including  Zenimal\footnote{https://zenimals.com/} and Moonbird\footnote{https://www.moonbird.life}. These devices guide users through meditation using vibration as a simple anchor point. In contrast, the current research frontier in tangible devices is advancing toward more sophisticated, immersive, and adaptive systems that integrate closed-loop sensing and feedback mechanisms ~\cite{terzimehicReviewAmpAnalysis2019}.

A key challenge for designing tangible mindfulness systems lies in identifying \emph{which stage or mechanism of mindfulness practice} the device intends to support, and how this intention translates into device's sensing, feedback, and how to evaluate for its success. From a design perspective, this raises several interrelated questions: \emph{what} aspects of the user's internal state should be sensed, \emph{how} accurately can this sensing be achieved within the constraints of wearability and compact form factors, and \emph{what} forms of feedback are most appropriate. While, in theory, increasing the number of sensors may improve the fidelity of internal state inference, such approaches are often limited by practical considerations of comfort, usability, and device intrusiveness. Feedback design introduces an additional layer of complexity. Designers must carefully consider \emph{what} type of feedback to provide, \emph{how} it should be delivered (e.g., auditory, haptic, or visual modalities), \emph{when} it should be presented to maximize effectiveness without disrupting the meditative process, and \emph{what} experiential or metaphorical meaning it conveys. This is particularly critical in mindfulness contexts, where the feedback mechanism effectively becomes the meditator's primary point of interaction, shaping their attentional and somatic experience throughout the practice.

While prior HCI work has proposed design frameworks for mindfulness in mobile contexts~\cite{salehzadehniksiratFrameworkInteractiveMindfulness2017}, there remains a limited understanding of how the embodied interaction paradigms of tangible devices can be fully leveraged. Consequently, many existing tangible mindfulness systems risk being reduced to simplistic guidance tools, rather than realizing their potential to facilitate richer, deeply embodied forms of practice. This gap motivates our central research question: \textbf{``How can tangible interaction be integrated in a theoretical framework explaining the relationship between somatic engagement and meditation experience?''}

We address this question through a systematic investigation grounded in existing HCI literature. We begin by analyzing the current landscape of tangible mindfulness devices through a scoping review (\hyperref[sec:scoping]{Section~\ref*{sec:scoping}}), structured around \citet{holzelHowDoesMindfulness2011}'s mechanisms of action of mindfulness: attention regulation, body awareness, emotion regulation, and change in perspective on the self. Specifically, as our first contribution, we examine which of these mechanisms each device engages with, and as our second contribution, we examine how these systems are evaluated in terms of achieving their intended mindfulness outcomes (\hyperref[sec:results]{Section~\ref*{sec:results}}).

Our findings are threefold. First, we identify substantial growth in this research area: while \citet{terzimehicReviewAmpAnalysis2019} report 34 studies which examined how technology can support mindfulness across VR, mobile applications, and tangible systems up to 2019, our search yielded 65 relevant papers on tangible systems alone. Second, existing work predominantly focuses on attentional regulation and body awareness, with comparatively less emphasis on emotion regulation and change in perspective on the self. Third, we found limited use of robust evaluation methodologies for determining whether these devices effectively achieve their intended mindfulness outcomes. We argue that these limitations stem, in part, from insufficient theoretical grounding in how tangible interaction can meaningfully support mindfulness practice.

Finally, to address the above challenge, we propose a theoretical framework with two complementary models for tangible mindfulness devices (\hyperref[sec:framework]{Section~\ref*{sec:framework}}). To support attention regulation, we introduce the \emph{Embodied Sensory Expansion (ESE)} model, extending \citet{vagoSelfawarenessSelfregulationSelftranscendence2012}'s concentrative practice model for Focused Attention Meditation (FAM), as our third contribution. Moreover, to address body-awareness, emotion regulation and changes in self-perspective, we propose the \emph{Embodied Sensory Anchoring (ESA)} model, building upon the principles of open monitoring receptive practice model proposed for Open Monitoring Meditation (OMM) by \citet{vagoSelfawarenessSelfregulationSelftranscendence2012} as our fourth contribution. Together, this framework aim to provide a theoretically grounded foundation for designing and evaluating tangible systems that fully realize the potential of embodied mindfulness interaction. 


\section{Scoping Review}
\label{sec:scoping}
This work adopts a scoping review methodology~\cite{munnSystematicReviewScoping2018}, which is well suited to examining and mapping the breadth of existing research on expansive and heterogeneous topics, such as tangible interfaces for mindfulness. 
We follow the methodological guidance from the Joanna Briggs Institute~\cite{petersUpdatedMethodologicalGuidance2020}, which extends the foundational framework proposed by~\citet{arkseyScopingStudiesMethodological2005} and aligns with the PRISMA Extension for Scoping Reviews~\cite{triccoPRISMAExtensionScoping2018}. The review process comprised five stages conducted by five researchers: (1) systematic database searching; (2) iterative, post hoc development and refinement of inclusion and exclusion criteria in line with established scoping review practices~\cite{arkseyScopingStudiesMethodological2005}; (3) study screening; and (4) data extraction and coding (5) consultation with key stakeholders (see \autoref{fig:PRISMA_flow}).

\subsection{Research Questions}

The aim of this research is to explore and structure the emerging field of tangible interaction for mindfulness. The following research questions guide the scoping review to provide both an overview of its human-centered foundations (RQ1) and an in-depth understanding of its tangible devices' evaluation characteristics (RQ2):

\begin{itemize}
    \item \textbf{RQ1:} How does HCI literature conceptualize the mechanisms of action of mindfulness within tangible interaction systems?
    \item \textbf{RQ2:} How does HCI literature evaluate the design of tangible interaction systems for mindfulness based on its intended goal?
\end{itemize}

\subsection{Mechanisms of Action of Mindfulness Meditation}
Recognizing the need to understand how mindfulness works from a conceptual, psychological and neuro-biological perspective, ~\citet{holzelHowDoesMindfulness2011} proposed its mechanisms of action. This includes four distinct mechanisms: (1) attention regulation, (2) body awareness, (3) emotion regulation: often split into reappraisal (3.1) and exposure, extinction, and reconsolidation (3.2), and (4) change in perspective on the self. Attention regulation forms the foundational skill, enabling sustained focus and conflict monitoring that prevents distraction and supports engagement with present-moment experience. Body awareness builds on this by enhancing sensitivity to internal bodily states, which are essential for recognizing and understanding emotional experiences. Emotion regulation operates through both reappraisal--reframing experiences in a more adaptive or meaningful way--and through exposure-based processes, where individuals observe emotions non-reactively, allowing extinction of habitual responses and re-consolidation of new patterns. Finally, a change in perspective on the self, often described as decentering, involves recognizing the transient nature of thoughts and emotions, reducing identification with them. Together, these mechanisms are highly interrelated: sustained attention facilitates awareness of bodily and emotional states, which in turn enables adaptive regulation and non-reactivity, ultimately fostering a shift in self-perspective.

\subsection{Positioning Within Existing Reviews}
Several existing reviews inform our work but are distinct from it. The closest is~\citet{terzimehicReviewAmpAnalysis2019}'s analysis of mindfulness in HCI, which identifies mobile applications, virtual reality, and tangible systems as primary technological modalities and charts the field according to mindfulness definitions and the role of technology; however, tangibles appear as just one category within a broader design space, and the review predates the field's substantial recent expansion.
~\citet{zhouTangibleAffectLiterature2024} offer a more recent perspective on tangible user interfaces for affect, yet their concern is affective interaction broadly rather than mindfulness as a distinct construct, with an emphasis on assessment and expression over intervention design. 
Similarly,~\citet{A_descriptive_haptic_system_design}'s review of affective haptic systems is limited to a single modality, whereas our scope encompasses tangible systems integrating multi-modal feedback. Finally, within the mindfulness literature,~\citet{sliwinski2017review} review interactive technologies for the cultivation of mindfulness, with a stronger emphasis on intervention outcomes. In contrast, our work adopts a design-oriented perspective.
Taken together, existing reviews either treat tangibles as one component of a wider design space or focus narrowly on affect or haptics; our work differentiates itself by providing a focused, up-to-date review of tangible systems for mindfulness specifically, charting the literature through the lens of~\citet{holzelHowDoesMindfulness2011}'s mechanisms of action.

\begin{figure}
    \centering
    \includegraphics[width=0.85\linewidth]{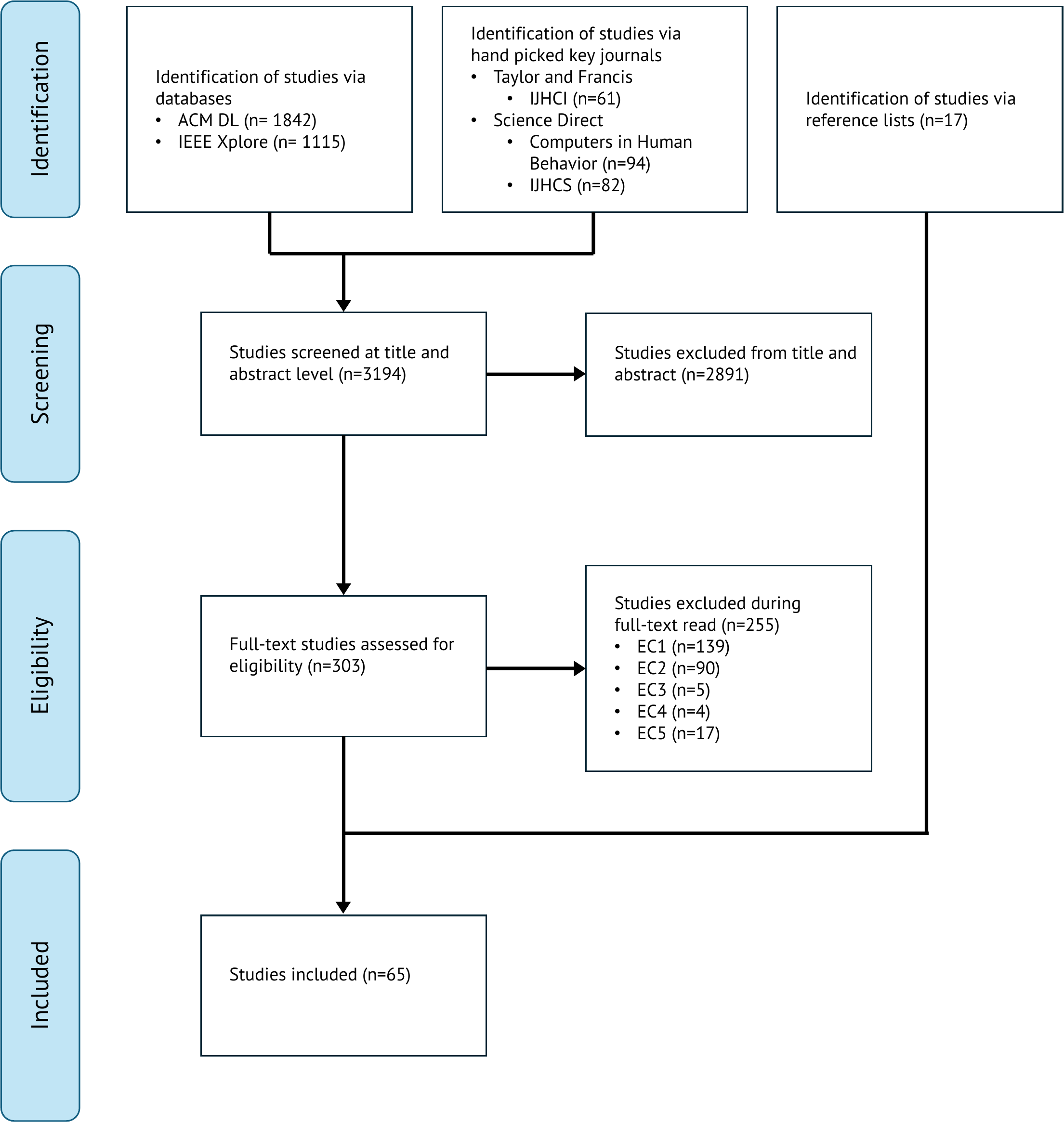}
    \caption{PRISMA~\cite{triccoPRISMAExtensionScoping2018} flow diagram of the scoping review process.}
    \label{fig:PRISMA_flow}
\end{figure}

\subsection{Search Strategy}

Following the recommendations of~\citet{arkseyScopingStudiesMethodological2005} \& ~\citet{petersUpdatedMethodologicalGuidance2020} for developing a comprehensive and iterative search strategy, we employed three complementary approaches to identify relevant literature. First, we conducted systematic searches across major publication venues in HCI that regularly feature work on tangible interfaces and mindfulness-related technologies. To ensure consistency with established practices in the field while maintaining a focused and manageable scope, we limited our primary database search to the ACM Digital Library and IEEE Xplore. These sources are widely recognized as central repositories for HCI research and have been similarly used in prior reviews, including~\citet{terzimehicReviewAmpAnalysis2019},~\citet{zhouTangibleAffectLiterature2024}, and~\citet{A_descriptive_haptic_system_design}.
Second, during our preliminary screening, we identified ScienceDirect (e.g., International Journal of Human-Computer Studies, Computers in Human Behavior) and Taylor \& Francis (e.g., International Journal of Human-Computer Interaction) as additional venues that frequently publish relevant work; these were subsequently incorporated into the search corpus. Third, we performed backward and forward citation chaining on studies identified, including screening the reference lists of eligible articles and relevant review papers to capture additional contributions (see \autoref{fig:PRISMA_flow}).

The search query was structured in two components and intentionally kept broad (following ~\citet{arkseyScopingStudiesMethodological2005}'s recommendations). The first component targeted mindfulness-related terminology, incorporating both \textit{mindfulness} and \textit{meditation}. Although the primary focus was on mindfulness, prior psychological literature indicates that repeated engagement in meditation practices contributes to the development of trait mindfulness~\cite{kikenStateTraitTrajectories2015}. 
The second component focused on interaction modalities, specifically \textit{tangible} technologies.
We first applied the query \emph{(``mindfulness'' OR ``meditation'') AND (``tangible'')} to full-text searches in ACM Digital Library. Then to improve coverage--particularly given the prevalence of wearable devices as a form of tangible technology--we subsequently expanded the query to include \textit{wearables}. Also we added keywords based on~\citet{Getting_a_grip_2006}'s framework to include the breadth of \textit{tangible interaction}, \textit{tangible user interfaces} and \textit{embodied interaction}. We leave out wording related to collaboration as we look into single person mindfulness.

After iterative refinement, the final search string was defined as:\\
\emph{(``mindfulness'' OR ``meditation'') AND (``tangible'' OR ``wearables'' OR ``embodied'' OR ``tangible interaction'' OR ``tangible manipulation'' OR ''embodied interaction'' OR ``hybrid physical systems'')}.

All searches were conducted as \textbf{full-text search} queries to maximize retrieval breadth and covered the complete range of available records up to \textit{12 May 2026}, the date of the final search. In total, 3194 papers were retrieved from the search process prior to screening (see \autoref{tab:paper_distribution})

\begin{table}[t]
\centering
\caption{The venues considered during first and second rounds of database searches}
\label{tab:venues}
\renewcommand{\arraystretch}{1.2}
\begin{tabularx}{\linewidth}{lX}
\hline
\textbf{Venue} & Number of Publications \\
\hline
\textbf{ACM Digital Library} & 1842 \\
\textbf{IEEE Xplorer}$^{*}$ & 1115\\
\textbf{Taylor and Francis} & \\
International Journal on Human Computing Interaction (IJHCI)$^{**}$ & 61 \\
\textbf{Science Direct}$^{+}$ &  \\
Computers in Human Behavior$^{1}$ & 94 \\
International Journal on Human Computing Interaction (IJHCS)$^{2}$ & 82 \\
\hline
\end{tabularx}
\vspace{0.5em}
\begin{minipage}{\linewidth}
\footnotesize
$^{*}$ In IEEE Xplore we used ``Full Text .AND. Metadata:'' for each \newline keyword in \textbf{advanced search $\rightarrow$ command search}.\newline
$^{**}$ In Taylor and Francis we searched \textbf{within} the IJHCI journal.\newline
$^{+}$ In Science Direct we used \textbf{advanced search} with separate queries for \newline
$^{1}$ International Journal of Human-Computer Studies as journal title \newline
$^{2}$ Computers in Human Behavior as journal title \newline
\end{minipage}
\label{tab:paper_distribution}
\end{table}



\subsection{Selection}
All identified citations were exported after the search (see \autoref{fig:PRISMA_flow}). Initially, the first author randomly sampled a subset of papers (n=10) and discussed them with two co-authors. Based on this collaborative review, preliminary exclusion criteria were developed. Then for the initial screening phase, the first author randomly selected 200 papers to use for assessing intercoder reliability to balance reliability and time~\cite{oconnorIntercoderReliabilityQualitative2020}. Three Authors reviewed each title and abstract using a four-point inclusion scale: Yes, Maybe Yes, Maybe No, and No. This resulted in a Krippendorff's alpha~\cite{krippendorffComputingKrippendorffsAlphareliability2011} of 0.595, indicating divergent agreement~\cite{marziKAlphaCalculatorKrippendorffs2024}. 
All papers with coder disagreements and \emph{Maybe} answers were then discussed in a post-coding meeting, which led to a following refined exclusion criteria.

\begin{enumerate}
    \item \textbf{EC1:} \textit{Lack of Explicit Relevance}: The study does not explicitly articulate usefulness for mindfulness or mindful meditation.
    \item \textbf{EC2:} \textit{Tangible System Absence}: The study does not present or evaluate a tangible, haptic, or wearable system (e.g. pure VR or Mobile Apps).
    \item \textbf{EC3:} \textit{Mindfulness Outcome Research}: The study investigates the psychological or physiological effects of mindfulness (e.g. effect of mindfulness on touch perception).
    \item \textbf{EC4:} \textit{Algorithmic Focus}: The study exclusively addresses internal computational methods or algorithmic processes (e.g., classification of meditative states) without involving a mindfulness intervention.
    \item \textbf{EC5:} \textit{Quality Control}: The record consists of non-primary research materials such as workshop calls, magazine articles, demonstrations, student design challenge or doctoral colloquium.
\end{enumerate}

A second round of coding was then conducted using a binary scale (Yes / No), which yielded a Krippendorff's alpha of 0.845, reflecting substantial agreement~\cite{marziKAlphaCalculatorKrippendorffs2024}. Given the acceptable intercoder reliability, first author proceeded to screen all remaining titles and abstracts against the inclusion and exclusion criteria, and identified 303 papers for full-text review.

The detailed rationale for each exclusion criterion with examples is outlined as follows.


\textbf{EC1} excludes studies that do not demonstrate clear relevance to mindfulness as intentional, present-moment attention, consistent with \cite{kabat_zinnMindfulnessbasedInterventionsContext2003}. Under \textbf{EC1}, \textit{Breathing-related systems} are excluded unless they explicitly support mindful attention; following the framing of breathing research in HCI \cite{prpaInhalingExhalingHow2020}, we exclude (i) systems relying on implicit or unconscious breathing regulation without attentional engagement (e.g., \cite{choiASpireClippableMobile2021, leeAmbientBreathUnobtrusiveJustintime2021, taborUnderstandingDesignEffectiveness2021, zhouEscentCoachWearable2025}), (ii) systems that explicitly position breathing guidance as non-meditative (e.g., \cite{Thomas_Dincel_Buchem_2025, yu2015breathe, yuViBreatheHeartRate2021, TechADeepBreath2026}), and (iii) soma design work centered on breathing without a mindfulness focus (e.g., \cite{Tsaknaki_Cotton_Karpashevich_Sanches_2021, karpashevichTouchingOurBreathing2022}). \textit{Non-mindful or passive emotion regulation systems} are excluded (e.g., \cite{costaEmotionCheckLeveragingBodily2016, costaEmotionCheckLeveragingBodily2016, costaBoostMeUpImprovingCognitive2019, papadopoulou2019affective, jainModulatingInteroceptiveSignals2023, umairExploringPersonalizedVibrotactile2021, zhaoAffectiveTouchImmediate2023}), as are \textit{self-reflection and personal informatics tools} that do not cultivate mindful awareness (e.g., \cite{MoodWings2013, thudtSelfreflectionPersonalPhysicalization2018, carpenter2019traekvejret, kuprijanovaChakraSuitExperimentalDirected2019, abtahiUnderstandingPhysicalPractices2020, yuCreativelySupportingMental2025}). We further exclude \textit{systems inspired by meditation but not intended for mindful practice}, such as exploratory or alternative-use prototypes (e.g., \cite{ladelfaDroneChiSomaesthetic2020, khotDesigningMicrobreaksUnpacking2022, aggarwalZenscapeEncouragingMicrobreaks2024, tenbhomerDesigningPersonalizedMovementbased2018}) and somaesthetic (e.g., inspired by Feldenkrais \cite{SomaestheticAppreciationDesign2016, stahlSomaMatBreathing2016}). Additionally, \textit{tools grounded in religious or ritual artifacts}, as well as \textit{conditioning and guided relaxation systems} (e.g., \cite{fraiettaTransientRelicsTemporal2020, markumDesignTechnosacredSpaces2025, linAromaCueScentToolkit2020, AltarNation2002}), are excluded. Finally, \textit{affective communication systems} (e.g., \cite{fooUserExpectationsMental2021, haynesJustBreathAway2024}) and \textit{systems primarily supporting rehabilitation, social or narrative experiences} rather than mindfulness practice (e.g., \cite{niedderer2007designing, nagargoje2012social, nunezpachecoTacitNarrativesSurfacing2017, freyBreezeSharingBiofeedback2018, SlowFloor2014, madapuranagarajMindfulnessbasedEmbodiedTangible2024a}) are excluded. \textbf{EC2} restricts the review to embodied, physical systems, excluding mobile (e.g. see~\cite{salehzadehniksiratFrameworkInteractiveMindfulness2017, kressbach2018breath}), VR (see~\cite{shamekhi2018breathe, rookHeartGardenVisualizing2025}), and purely cloud-based approaches. These modalities have been extensively covered in prior reviews (e.g., see~\cite{terzimehicReviewAmpAnalysis2019}) and are therefore outside the scope of this work. 
\textbf{EC3} removes studies that focus exclusively on outcomes of mindfulness interventions (e.g., clinical outcomes and enhance task performance~\cite{vasudevanMindTouchEffectMindfulness2023}), as this body of work is well established within psychology and contemplative science (e.g.,~\cite{khouryMindfulnessbasedTherapyComprehensive2013}), and does not directly contribute to the design of interactive systems. 
\textbf{EC4} excludes studies that focus solely on algorithmic development or sensing techniques without an explicit interaction design contribution (e.g. \cite{Hao_Bi_Xing_Chan_Tu_2017}. Such work is more appropriately situated within the affective computing literature (e.g., see~\cite{zhouTangibleAffectLiterature2024}). \textbf{EC5} excludes records that consist of non-primary research materials such as workshop calls, magazine articles, demonstrations, student design challenge or doctoral colloquium (e.g. see \cite{Tag_Goto_Minamizawa_Mannschreck_Fushimi_Kunze_2017, van_Rheden_Liu_Luo_Elvitigala_Mueller_2025, Tao_2025, CraftingWearables2013, stahlSomaMatBreathing2016}).

The full-text of the selected results was then carefully evaluated using the same inclusion and exclusion criteria and final subset of 48 papers were selected from this process. Next by citation chaining of the 48 papers, additional 17 papers were included to the review.

For methodological constraints in screening process see~\hyperref[sec:Appendix_1]{Appendix I}.


\subsection{Data Extraction and Analysis}
The final subset of records comprised 65 papers. For data extraction,  we employed a structured approach using a data charting form, as described in the methodology for scoping reviews by~\citet{petersUpdatedMethodologicalGuidance2020}, which builds on~\citet{arkseyScopingStudiesMethodological2005}. After reading 35 papers to decide on inclusion, the first author developed a preliminary data charting form informed by the two research questions. This preliminary data charting form included the categories: general information, research objectives and contributions, user evaluations/empirical studies, mindfulness information, and device's technical information. To coordinate and validate the data extraction procedure, three authors selected four representative studies--two qualitative~\cite{dublinWalkingMeditationMat2025a, liCodesigningMagicMachines2023} and two quantitative~\cite{farrallManifestingBreathEmpirical2023a, gemiciogluBreathePulsePeripheralGuided2024}--and independently extracted data using the preliminary charting form. The team then compared results, discussed the procedure, and refined the protocol. Two meetings were held during this process to evaluate the suitability of the charting form, which was iteratively updated as the project progressed. Following this calibration, the first author completed data extraction for all the papers in the review set. The final chart, including all subcategories and attributes, is provided in~\hyperref[appendix:charting_protocol]{Appendix III}

To address RQ1, we examined how mindfulness is represented in the literature based on four dimensions. First, we analyzed the \textbf{theoretical grounding} of each work--whether it adheres to established clinical paradigms such as MBSR or adopts a more implicit or bespoke definition of mindfulness. Second, we investigated how mindfulness is \textbf{embodied and enacted} within the system by examining the physical postures or movement modalities supported following \citet{terzimehicReviewAmpAnalysis2019}'s role of technology. Third we looked into the devices \textbf{sensing and feedback modalities} used. We used these information along with other author provided information to identify the underlying \textbf{psychological and neuro-cognitive processes} targeted based on the conceptual and neural frameworks as proposed by~\citet{holzelHowDoesMindfulness2011}.

To address RQ2, we employed an inductive approach to examine how tangible interaction systems are evaluated within the HCI literature. Our team systematically logged information regarding study design, evaluation metrics, and reported outcomes in relation to each system's stated objectives. 
Subsequently, we organized these evaluation approaches into coherent thematic categories under psychological, physiological and usability themes to facilitate clear and systematic presentation.

\subsection{Consultation with stakeholders}

As a final stage of the scoping review, we conducted a consultation with key stakeholders, which~\citet{arkseyScopingStudiesMethodological2005} argue can add value to both the literature review process and its overall findings. 
Hence, we designed and conducted a group consultation with eight participants, which closely aligns with the typical consultation size reported in prior scoping reviews (median = 9)~\cite{buus2022arksey}, involving mindfulness practitioners, mindfulness-based stress reduction (MBSR) participants, and HCI designers. To ensure methodological rigour across findings, we provide a detailed account of the consultation process in a dedicated section (see \hyperref[sec:consultation]{Section~\ref*{sec:consultation}}).

\section{Results}
\label{sec:results}

\begin{figure}
    \centering
    \includegraphics[width=0.8\linewidth]{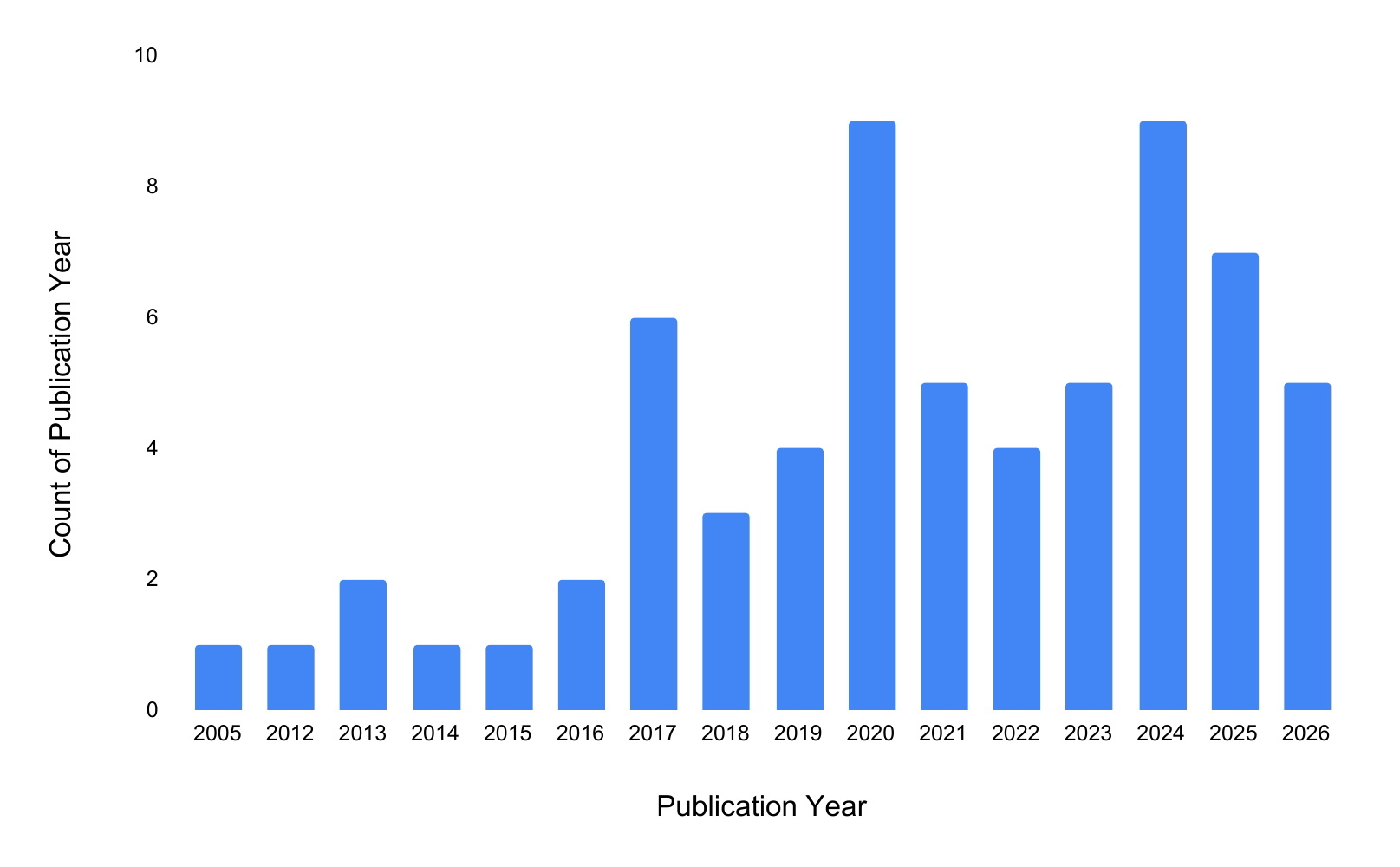}
    \caption{Overview of publication trends across the final subset of papers, showing the number of publications by year.}
    \Description{The chart shows yearly publication counts, with publications generally increasing over time and peaking in 2024}
    \label{fig:charted_year}
\end{figure}

The final subset of records comprised 65 papers (see \autoref{tab:overview_of_all_articles}), including 48 conference papers, 16 journal articles, and 1 industrial article published by Philips. 
It should be noted that the count for 2026 includes only papers published up to 12 May 2026; therefore, the observed trend may continue upward, suggesting growing interest within the HCI community. An overview of publication trends by year~\autoref{fig:charted_year}.

\subsection{\textbf{RQ1}: How is mindfulness operationalized in HCI tangible device research, particularly with respect to the psychological mechanisms being targeted?}

To answer our first research question, we organized the literature into distinct substreams based on the four mechanisms of action of mindfulness (by \citet{holzelHowDoesMindfulness2011}) that are supported by the devices themselves. Through this deductive coding process, five main sub-streams emerged.

\begin{itemize}
    \item \textbf{Attention Regulation (n=28):} Papers that present systems or approaches supporting the sustained focus of attention on a chosen object and the redirection of attention when distractions occur.
    \item \textbf{Body Awareness (n=20):} Papers that present work supporting attention to internal experiences, including sensory awareness of breathing, emotions, and other bodily sensations.
    \item \textbf{Emotion Regulation (n=5):} Papers that present approaches enabling users to relate to ongoing emotional reactions in a different way (non-judgmentally, with acceptance)
    \item \textbf{Change in Perspective on the Self (n=5):} Papers that present work facilitating detachment and decentering from identification with a fixed or static sense of self.
    \item \textbf{Meta papers (n=7):} Papers that present work overarching one or more mechanisms.
\end{itemize}

In the following, we summarize the findings for each mechanism of action of mindfulness. For each, we begin with a brief conceptual background~\cite{holzelHowDoesMindfulness2011} and then review the corresponding literature. An overview of all mechanisms and their associated papers is provided in \autoref{tab:overview_of_all_articles}.

\begin{table*}
    \centering
    \begin{tabular}{p{5cm} p{9cm}}
       Mindfulness Mechanism  & Paper\\
       \toprule
       
       Attention Regulation & \cite{feijs2005design}, \cite{vidyarthi2012sonic}, \cite{MindPool2013}, \cite{vidyarthiInteractivelyMediatingExperiences2014}, \cite{sas2015meditaid}, \cite{Engagement_through_Embodiment_16}, \cite{macik2017breathinga}, \cite{paredesJustBreatheIncar2018}, \cite{paredes2017evaluating}, \cite{pryssPersonalizedSensorSupport2018}, \cite{Mediscape2020}, \cite{fooSoftRoboticCompression2020}, \cite{miriPIVPlacementPattern2020}, \cite{miriEvaluatingPersonalizableInconspicuous2020}, \cite{daudenroquetInteroceptiveInteractionEmbodied2021a}, \cite{khotSWANDesigningCompanion2020}, \cite{FirstPersonWalking2021}, \cite{choiDesignEvaluationClippable2022}, \cite{tanMindfulMomentsExploring2023}, \cite{ezerSomaestheticMeditationWearable2024a}, \cite{gemiciogluBreathePulsePeripheralGuided2024}, \cite{wang2024design}, \cite{dublinJourneyInwardSomaesthetic2024}, \cite{choMindfulTouchMidair2025}, \cite{chenLivingBentoHeartbeatDriven2025}, \cite{dublinWalkingMeditationMat2025a}, \cite{BreathingInward2026}, \cite{Samten2026}\\
       
       Body Awareness & \cite{thiemeDesignPromoteMindfulness2013a}, \cite{aslanHoldMyHeart2016}, \cite{tenbhomerDesigningPersonalizedMovementbased2018}, \cite{semertzidisUnderstandingDesignPositive2019}, \cite{aslanPiHeartsResonatingExperiences2020}, \cite{chinarevaLotusMediatingMindful2020}, \cite{choiAmbienBeatWristwornMobile2020a}, \cite{jungExploringAwarenessBreathing2021a}, \cite{sabinsonPlantHumanEmbodiedBiofeedback2021}, \cite{EtherealPhenomena2022}, \cite{KhongKhro2022}, \cite{kuDisImmersionMindfulness2023}, \cite{farrallManifestingBreathEmpirical2023a}, \cite{SMArtBracelet2023}, \cite{AmbientPlantforMeditation2024}, \cite{ConsciousOrUnconsciousMeditation2025}, \cite{hyunVibroCushionDesignInclusive2025}, \cite{tanRunMeAdaptiveSound2025}, \cite{EncouragingBreath2026}, \cite{EphemeralBreath2026}\\
       
       Emotion Regulation & \cite{seolDropBeatVirtual2017}, \cite{Sprite_Catcher_2017}, \cite{huttonReMiNDImprovingEmotional2019}, \cite{cochraneBreathingScarfUsing2022},  \cite{huangCoralMorphArtistic2025}\\
       
       Change in Perspective on the Self &   \cite{rooInnerGardenConnecting2017a}, \cite{zhuDesigningDigitalMindfulness2017}, \cite{vianelloTANGAEONTangibleInteraction2019}, \cite{wangReflectingSoloDining2025}, \cite{mahDesigningRitualInteraction2020}\\
       
       Meta papers & \cite{terzimehicReviewAmpAnalysis2019}, \cite{daudenroquetBodyMattersExploration2020}, \cite{liCodesigningMagicMachines2023}, \cite{liMeditationUnderstandingEveryday2024a}, \cite{markumMediatingSacredConfiguring2024}, \cite{staabCanYouBe2024}, \cite{zhouTangibleAffectLiterature2024} \\
       
       \bottomrule
    \end{tabular}
    \caption{Overview of all articles included within this review, across the different mechanisms of action proposed by~\citet{holzelHowDoesMindfulness2011}}
    \label{tab:overview_of_all_articles}
\end{table*}

\subsection*{Attention Regulation}
We found 28 papers (43\%) related to supporting attention regulation during mindfulness activities. According to~\citet{holzelHowDoesMindfulness2011}, attention regulation refers to the capacity to sustain attention on a chosen object (e.g., breath, emotions, mantra) and, upon distraction, intentionally reorient attention back to that object. Within these papers, we identified four main design strategies through open coding.

\paragraph*{3.2.1.1 Externalized attention anchors (9/28)}
\label{para:attention_reg_externalized_attention_anchors}

The design strategy focuses on externalizing the attentional object itself, most commonly through breathing-based entrainment \cite{macik2017breathinga, miriPIVPlacementPattern2020, miriEvaluatingPersonalizableInconspicuous2020, choiDesignEvaluationClippable2022, gemiciogluBreathePulsePeripheralGuided2024, wang2024design, tanMindfulMomentsExploring2023}. A shared characteristic across these systems is the use of open-loop breathing pacing, which creates a persistent and temporally structured attentional cue. In contrast, their primary divergence lies in the modality through which this cue is delivered. For instance, \citet{tanMindfulMomentsExploring2023} propose \textbf{Mindful Moments}, an optical head-mounted display (OMHD) that provides visual cues for inhalation and exhalation, exemplifying a purely visual modality. In contrast, \cite{macik2017breathinga} introduce a \textbf{handheld tangible device} with an internal electromechanical system that produces rhythmic inflation and deflation, enabling haptic interaction that mimics a breathing pattern. In \cite{miriPIVPlacementPattern2020}, \cite{miriEvaluatingPersonalizableInconspicuous2020} authors explore the pattern, placement, shape of the feedback for guided mindful slow breathing through vibration. Similarly, \citet{wang2024design} present \textit{Mysa}, a \textbf{haptic garment} equipped with vibration motors that guide breathing through a structured pattern. While it shares the same open-loop pacing principle, it diverges through full-body wearable haptics. In another haptic-based approach, \citet{choiDesignEvaluationClippable2022} develop a wearable device that delivers pneumatic feedback via soft inflatable actuators, where the goal breathing rate (GBR) is dynamically set to 70\% of the user's mean breathing rate (MBR), introducing a semi-personalized variation within the same design paradigm. Another example, BreathPulse \cite{gemiciogluBreathePulsePeripheralGuided2024}, provides controlled airflow for breathing guidance through the device mounted on the back of a laptop. This approach shifts the modality from on-body feedback to environmental augmentation, while still maintaining the same underlying open-loop entrainment structure. Qualitative findings indicate that this system promotes mindfulness. Despite their effectiveness in stabilizing attention, these systems largely emphasize externally driven attentional control. Beyond wearable and ambient systems, similar design principles extend into automotive contexts.

A related but distinct sub-strategy shifts the attentional anchor from respiration to \textbf{somatic sensation}, while still externalizing attention through structured temporal cues \cite{fooSoftRoboticCompression2020, choMindfulTouchMidair2025}. For example, \citet{fooSoftRoboticCompression2020} present a \textbf{soft robotic compression garment} that applies rhythmic compression and warmth to the shoulders. Unlike breathing-based systems, this design does not explicitly encode inhale-exhale timing, but instead grounds attention in bodily sensation and affective touch, leveraging cyclic compression-relaxation patterns to reduce mind wandering. Similarly, \citet{choMindfulTouchMidair2025} explore \textbf{mid-air haptics} as a contactless, somatosensory modality layered onto audio-guided meditation. Rather than enforcing a fixed breathing rhythm, the system delivers subtle tactile stimuli on the palm that users may optionally align with their breathing. This introduces a looser coupling between external cues and respiratory processes, enabling a more flexible and user-driven attentional regulation. Notably, findings suggest that such stimuli may initially disrupt attention but subsequently facilitate mindfulness through processes such as embodied grounding and attentional anchoring, highlighting a more dynamic and adaptive role of external cues.

\paragraph*{3.2.1.2 Closed-Loop Systems} (10/28) 
\label{para:attention_reg_closed_loop_systems}
We found ten papers which provide closed-loop systems as the design strategy and operationalize attention regulation as a dynamic sensing–feedback cycle~\cite{vidyarthi2012sonic, MindPool2013, vidyarthiInteractivelyMediatingExperiences2014, sas2015meditaid, paredesJustBreatheIncar2018, Mediscape2020, FirstPersonWalking2021, daudenroquetInteroceptiveInteractionEmbodied2021a, pryssPersonalizedSensorSupport2018, khotSWANDesigningCompanion2020}. Within the the biggest sub-strategy was \textbf{physiological and neuro-based closed-loop systems} which directly infer internal attentional states through biosignals and adapt feedback accordingly. For example, using physiological sensing, \citet{vidyarthi2012sonic} proposes Sonic Cradle, a chamber of complete darkness which uses real-time sonic feedback to make subtle respiratory variations audible, thereby externalizing breath dynamics through sound (extension of this work in \cite{vidyarthiInteractivelyMediatingExperiences2014}). The system is intended to help with users ``meta-awarness''. In contrast using neurological sensing, \citet{MindPool2013}'s MindPool, the user's mental states are detected using EEG and the users are given ambiguous feedback through magnetically reactive liquid to maintain the engaging interaction. In a similar EEG sensing, \citet{sas2015meditaid}'s \textit{meditAid}, uses neurofeedback to detect fluctuations in attentional focus and dynamically modulates auditory entrainment to guide users back to a focused state. \citet{Mediscape2020} extends this idea by focusing on qualitative experience during the meditation. And in their follow up work~\cite{FirstPersonWalking2021}, they extend felt-sense to walking meditation. In contrast, \citet{paredesJustBreatheIncar2018} propose a system that detects stress level and provide closed-loop breathing intervention in a driving context, where haptic guidance continuously structures inhalation and exhalation cycles to regulate autonomic arousal. In similar work in WarmMind~\cite{daudenroquetInteroceptiveInteractionEmbodied2021a}, the author use embodied metaphor theory, to explain how discrete thermal stimulation mapped to meditation states (e.g., mindfulness, mind-wandering)

In contrast the other sub-strategy, \textbf{behavioral closed-loop systems}, infer attentional state indirectly through observable actions and intervene when deviations from desired behavior are detected. Rather than measuring internal physiological signals, these systems rely on behavioral proxies of attention such as movement patterns, interaction habits, or task engagement. For instance, \citet{pryssPersonalizedSensorSupport2018} monitor walking speed and provide haptic feedback when users deviate from a predefined ``mindful'' pace. Here, attentional lapse is operationalized as behavioral drift, and correction is triggered when this deviation exceeds a threshold. Similarly, \citet{khotSWANDesigningCompanion2020} detect attentional distraction during eating (e.g., engagement with digital devices) and intervene through a mechanized spoon that disrupts the behavior.

\paragraph*{3.2.1.3 Somaesthetic Attention Anchors (4/28)}  
\label{para:attention_reg_somaesthetic_attention_anchors}

We identify another distinct strategy that externalize attention through continuous somaesthetic stimulation~\cite{ezerSomaestheticMeditationWearable2024a, dublinJourneyInwardSomaesthetic2024, dublinWalkingMeditationMat2025a, BreathingInward2026}, rather than temporally structured guidance (e.g., breathing rhythms) or sensing-driven adaptation. Unlike externalized entrainment approaches, these systems do not prescribe when or how attention should shift. A key example is the \textbf{somaesthetic meditation wearable} by \citet{ezerSomaestheticMeditationWearable2024a}, which delivers localized thermal stimulation to different body locations during meditation. The system operates in an open-loop manner, producing gradual heating and cooling cycles, but does not encode a target rhythm (e.g., inhale–exhale pacing). Instead, users are encouraged to reposition the device and explore how warmth interacts with their ongoing bodily awareness. Findings show that the warmth is often perceived as an internal sensation that ``pulls'' attention toward the body, facilitating grounding and introspection. In similar work, \citet{BreathingInward2026} uses targeted-heat wearable placed on the abdomen to guide diaphragmatic breathing via thermal pulses. The results reports increased state mindfulness and higher interoception. A related instantiation is the \textbf{walking meditation mat}, which applies distributed thermal patterns during slow walking meditation~\cite{dublinJourneyInwardSomaesthetic2024, dublinWalkingMeditationMat2025a}. Unlike breathing-based systems, the mat does not enforce a fixed temporal alignment with respiration. Instead, it provides spatially distributed and slowly evolving heat stimuli that guide awareness across while walking. This creates a form of attentional anchoring grounded in bodily exploration and spatial progression, rather than rhythmic entrainment.

\paragraph*{3.2.1.4 Task-Coupled Attentional Constraints (5/28)} 
\label{para:attention_reg_task_coupled}

We found five which embeds attention regulation directly within ongoing physical activity, such that sustained attention becomes a prerequisite for successful interaction~\cite{feijs2005design, Engagement_through_Embodiment_16, paredes2017evaluating, chenLivingBentoHeartbeatDriven2025, Samten2026}. Rather than guiding attention through externally imposed cues (e.g., rhythmic breathing signals) or correcting deviations via sensing-driven feedback, these systems operationalize attention as an emergent property of \textit{task engagement and sensorimotor coordination}. For example, \citet{feijs2005design} present \textit{MindSpheres}, a tangible system consisting of two handheld spheres that users are encouraged to twirl smoothly within the palm. The interaction requires fine motor control and continuous coordination, effectively binding attention to the dynamics of the task. The system analyzes movement in real time and provides light and vibrotactile feedback that becomes increasingly structured and harmonious as the user achieves smoother motion. Importantly, this feedback does not prescribe a target state (e.g., a breathing rhythm), but instead reflects the quality of ongoing interaction, reinforcing a state of sustained, flow-like engagement. In similar work \citet{Samten2026} propose Samten, three modular zen-inspired stones that requires stacking on top. The device provide vibration, light, and stone movement as feedback when attention drifts. If distraction persists, the stack physically collapses, serving as a signal of disrupted workflow. A related instantiation of this strategy emphasizes continuous embodied control and expressive coupling rather than discrete skill refinement. \citet{Engagement_through_Embodiment_16} redesign a common kitchen appliance--the blender--by with mindful interaction designed specifically around embodied interaction using redesigned turning, pushing, and pulling mechanisms. These interaction styles require users to continuously modulate their bodily input (e.g., precise movements using fingers) in order to sustain the system's operation, effectively coupling attention to the unfolding interaction. \citet{paredes2017evaluating} explore \textbf{in-car vibrotactile guidance} through a seat embedded with an array of actuators that deliver rhythmic patterns to guide breathing and body movements. These cues are temporally structured (e.g., inhale--hold--exhale sequences) and operate in an open-loop manner, functioning as external prompts rather than adaptive feedback. While consistent with the broader entrainment paradigm, this work is notable in explicitly proposing future integration of sensing (e.g., breathing rate, posture, physiological signals) to enable closed-loop feedback and adaptive guidance. Considering the act of mindful eating, Living Bento~\cite{chenLivingBentoHeartbeatDriven2025} maps heartbeat data captured through chopsticks to LED illumination beneath translucent noodles, integrating cardiac awareness into eating without requiring a separate mindfulness exercise. The study found the illumination improved bite-to-bite attentiveness during the eating process.


\subsection*{Body Awareness}
We found 20 papers (30\%) related to supporting body-awareness during mindfulness activities. According to~\citet{holzelHowDoesMindfulness2011}, body awareness in mindfulness refers to attending to internal bodily experiences, including breathing, emotions, and other somatic sensations. It is closely linked to interoception, the perception of internal physiological states, and plays a critical role in enhancing awareness of emotional processes \cite{holzelHowDoesMindfulness2011}. Within these papers, we identified three main design strategies through open coding.

\paragraph*{3.2.2.1 Embodied Amplification (7/20)} 
\label{para:body_awarness_embodied_amp}

This strategy focuses on amplifying internal bodily processes by rendering them directly perceivable through somatic feedback~\cite{aslanHoldMyHeart2016, aslanPiHeartsResonatingExperiences2020, choiAmbienBeatWristwornMobile2020a, jungExploringAwarenessBreathing2021a, farrallManifestingBreathEmpirical2023a, SMArtBracelet2023, EncouragingBreath2026}. Analogous to \textit{externalized attention anchors} in attention regulation, these systems externalize an otherwise internal target (e.g., heartbeat, breath), but differ in that the goal is not to stabilize attention through temporal entrainment, but to increase perceptual access to bodily signals. 

Several systems focus on cardiac and breathing awareness. \citet{aslanHoldMyHeart2016} introduce two artefacts one that renders heartbeat and second, stuffed animal that breathing perceptible through haptic feedback, explicitly aiming to foreground internal bodily processes through somaesthetic introspection. In their later work, PiHearts~\cite{aslanPiHeartsResonatingExperiences2020}, they further investigated the handheld silicone heart that physically beats in synchrony with the user's heart via sensor-actuator coupling. 
Similar to the cardiac awareness, ambienBeat~\cite{choiAmbienBeatWristwornMobile2020a} takes a more wearable approach by rendering heart rate as subtle wrist-based tactile pulses. Compared with PiHearts, which invites focal interaction with a tangible object, ambienBeat supports more continuous and low-effort awareness through a peripheral, on-body form factor. Shifting focus to breathing, \citet{jungExploringAwarenessBreathing2021a} use a wearable torso garment with pneumatically actuated pads that guide breathing through deep touch pressure on the back. This makes respiration spatially and physically salient, allowing users to feel breathing as localized pressure changes across different body regions. Similarly, \textit{Manifesting Breath} by \citet{farrallManifestingBreathEmpirical2023a} transform breathing into tangible movement through a pneumatic artefact that expands and contracts with the user's respiration. However, compared to \citet{jungExploringAwarenessBreathing2021a}'s system that maps breathing back onto the torso and supports anatomical awareness of ribs, diaphragm, and back, \textit{Manifesting Breath} creates a kinaesthetic analogue of lung movement through an object held in the hands (their extension of long term effect with updated device \cite{EncouragingBreath2026}). 

\paragraph*{3.2.2.2 Externalized Biofeedback Representations (8/20)} 
\label{para:body_awarness_externalized_biofeedback}

The largest pool of papers propose designs that externalize internal bodily processes into \textit{perceptual representations outside the body}~\cite{ thiemeDesignPromoteMindfulness2013a, semertzidisUnderstandingDesignPositive2019, sabinsonPlantHumanEmbodiedBiofeedback2021, EtherealPhenomena2022, KhongKhro2022, AmbientPlantforMeditation2024, tanRunMeAdaptiveSound2025, EphemeralBreath2026}. Unlike embodied amplification, where feedback is re-integrated and felt as a bodily sensation, these systems maintain the representation in the environment, positioning internal states as \textit{objects to be observed and interpreted}. 

A shared characteristic across these systems is the translation of physiological signals into visual, auditory, kinetic, or ambient media. Their primary divergence lies in modality, scale, and the degree to which the representation demands focal versus peripheral attention. The Mindfulness Sphere~\cite{thiemeDesignPromoteMindfulness2013a} similarly anchors attention in an internal signal, but uses visual and tactile feedback to represent the heartbeat. While Vidyarthi et al. emphasize auditory amplification of breathing, the Mindfulness Sphere focuses on cardiac awareness through a tangible object. In Etheral Phenomena~\cite{EtherealPhenomena2022}, authors present an artwork, which captures thoracic and abdominal breathing and translate these physiological signals into responsive visual and auditory feedback. In contrast in Khong Khro~\cite{KhongKhro2022}, authors present an artwork which supports focused attention meditation systems by externalizing EEG biofeedback. A similar system that use mmWave radar for breathing regulation is proposed in \cite{EphemeralBreath2026}. 

More immersive instantiations extend this principle further. Inter-Dream \cite{semertzidisUnderstandingDesignPositive2019} uses EEG activity to modulate visual projections and VR imagery, allowing users to experience changes in brain activity as dynamic visual phenomena. pheB \cite{sabinsonPlantHumanEmbodiedBiofeedback2021} transforms heart rate and plant bio-signals into soft robotic inflation patterns, creating a nature-inspired biofeedback loop that users can see and feel.

More recent systems embed externalized biofeedback into everyday or aesthetic contexts. For example, \citet{AmbientPlantforMeditation2024} propose a interactive ambient plant which uses light as a breathing guidance and a water release mechanism which triggers if user's pulse rises below a threshold. And the device waters the ambient plant as a positive reward for maintaining the breathing and heart-rate under a threshold. RunMe~\cite{tanRunMeAdaptiveSound2025} applies the same principle in movement practice by using smartwatch data, including heart rate and cadence, to adapt auditory feedback during mindful running. Compared with Living Bento and Coral Morph, which primarily support reflective awareness of physiological state, RunMe also uses biofeedback to regulate ongoing bodily activity by guiding rhythm, exertion, and cadence.

\paragraph*{3.2.2.3 Scaffolded Somatic Attention (5/20)}
\label{para:body_awarness_scaffolded_somatic_attention}

In contrast to embodied amplification, which primarily increases signal salience, a third design strategy supports body awareness by structuring how attention is directed toward the body. These systems guide the \textit{trajectory of attention} across bodily regions or processes. This parallels \textit{closed-loop systems} and \textit{somaesthetic anchors} in attention regulation, where attention is guided either dynamically or through structured cues ~\cite{kuDisImmersionMindfulness2023, tenbhomerDesigningPersonalizedMovementbased2018, chinarevaLotusMediatingMindful2020, ConsciousOrUnconsciousMeditation2025, hyunVibroCushionDesignInclusive2025}.

A shared characteristic across these systems is the use of spatial, rhythmic, or instructional cues to reduce the effort required to locate and sustain attention on bodily sensations. Their primary divergence lies in the modality and degree of coupling between guidance and bodily processes. 
For instance, \citet{kuDisImmersionMindfulness2023} use spatialized audio emerging from anatomically corresponding locations on a mat, reducing the cognitive mapping required during body-scan practices. While both systems guide spatial attention, they differ in modality (thermal vs.\ auditory) and explicitness of instruction. In a similar guidance, \cite{ConsciousOrUnconsciousMeditation2025} use haptic cues to direct attention towards `chakras' in the body during VR meditation. In contrast, the Body Cushion by \citet{hyunVibroCushionDesignInclusive2025} has a touch-based interface which enables meditation instructors to guide body scan meditation in the meditators wearable by providing targeted vibration.

In movement-based contexts, scaffolding shifts toward proprioceptive and muscular awareness. \citet{tenbhomerDesigningPersonalizedMovementbased2018} use EMG sensing to detect muscle activation and provide multimodal feedback during yoga, operationalizing body awareness as correct engagement of specific muscle groups. This contrasts with receptive body-scan systems by emphasizing active, goal-directed awareness. The Lotus system~\cite{chinarevaLotusMediatingMindful2020} further diverges by scaffolding awareness through rhythmic breathing guidance driven by sensed stress levels, positioning awareness as emerging from synchronization with an external pacing structure.

\subsection*{Emotion Regulation}
\label{sec:emotion_regulation}

We found 5 papers (7\%) supporting emotion regulation during mindfulness activities. According to~\citet{holzelHowDoesMindfulness2011}, emotion regulation in mindfulness involves modifying one's relationship to emotional experience, either through reinterpretation (reappraisal) or through sustained, nonreactive exposure to affective states (extinction and reconsolidation). Within mindfulness, reappraisal involves a shift in how emotions are interpreted or related to, rather than direct suppression or avoidance. We found the five papers to offer four distinctive strategies for reappraisal. The Drop the beat system by \citet{seolDropBeatVirtual2017} externalizes interoceptive signals and transforms them into perceivable and manipulable representations. Using VR, heart-rate sensing, and haptic feedback, users are exposed to panic-inducing scenarios and subsequently interact with a virtual representation of their own heartbeat. By allowing users to ``hold'' and observe their heart as a tangible object synchronized with physiological data, the system reframes a typically threatening internal sensation into something concrete and controllable. 

A second sub-strategy supports reappraisal through structured reflection and emotional labeling. ReMiND~\cite{huttonReMiNDImprovingEmotional2019} proposes a device that detects physiological arousal and triggers haptic prompts that encourage users to pause and reflect, complemented by a companion application for emotion labeling, journaling, and contextual annotation. In contrast to immersive systems like Drop the Beat, which emphasize in-the-moment perceptual reinterpretation, ReMiND operates through retrospective structuring of experience. By converting raw affective responses into labeled and contextualized representations, it enables users to reorganize how emotional events are understood over time. The third variation combines physiological regulation with reflective practices to indirectly support reappraisal. The Breathing Scarf~\cite{cochraneBreathingScarfUsing2022} integrates biofeedback-driven breathing guidance with journaling and body-mapping exercises. While the wearable component primarily stabilizes attention and reduces physiological arousal, the surrounding reflective practices create conditions for reappraisal. Compared to ReMiND's discrete event-based reflection, this approach situates reappraisal within an ongoing, embodied process of sense-making. In similar work, Coral Morph~\cite{huangCoralMorphArtistic2025} uses heart rate to drive pneumatic movement and LED changes in an installation, producing a multisensory representation of physiological state. The installation support reappraisal through sense-making.

We did not identify tangible interactive systems that explicitly operationalize \textbf{exposure, extinction, or reconsolidation} as a primary design strategy as proposed by~\cite{holzelHowDoesMindfulness2011}. Such device should employ a mechanism involving sustained, nonreactive engagement with emotional experience, supporting habitual stimulus–response patterns. However, we found partial implementation of the mechanism in Sprite Catcher~\cite{Sprite_Catcher_2017}. In the proposed design author suggest the device is designed to disrupt rumination happening at the moment by redirecting attention by `catching' the ruminative thoughts using the device. However, the self-reflection happens later--not real-time--as proposed ``during the session [counseling], Laura [..] talks about each [emotions] with her counselor in turn, reflecting upon the experiences they represent.'' 

\subsection*{Change in Perspective on the Self} 
\label{sec:change_in_perspective}

We found 5 papers (7\%) related to supporting change in perspective on the self. According to \citet{holzelHowDoesMindfulness2011}, Change in perspective on the self refers to a shift from identifying with a fixed sense of self toward experiencing thoughts, emotions, and sensations as transient events. Often described as \textit{decentering}, this process involves developing an observer perspective and meta-awareness of experience~\cite{holzelHowDoesMindfulness2011}.

A prominent approach involves metaphorical and environmental externalization. The Inner Garden system~\cite{ rooInnerGardenConnecting2017a} transforms physiological and affective signals into a miniature, evolving ecosystem. Internal states such as breathing and stress are mapped onto environmental dynamics including water movement, vegetation growth, and ecological balance. This creates a persistent, spatialized representation that can be observed over time. Compared to direct biofeedback systems, which increase perceptual salience, Inner Garden introduces abstraction that supports reflection. By rendering internal processes as an external environment, the system enables users to adopt a third-person observational stance.

A second group of systems supports decentering through tangible interaction with representations of thoughts or behaviors. TANGAEON~\cite{vianelloTANGAEONTangibleInteraction2019} allows users to input thoughts that are visually represented beneath a water-filled container. Through physical interaction, users generate waves that gradually dissolve these representations, reinforcing their impermanent nature. Similarly, the solo dining system by~\citet{wangReflectingSoloDining2025} externalizes habitual behavior into physical data artifacts that users assemble and annotate. While TANGAEON focuses on momentary cognitive content, this system emphasizes temporal aggregation and retrospective reflection. Both approaches convert implicit internal processes into manipulable objects, enabling a shift from immersion to observation.

A third strategy reduces self-referential processing by minimizing goal-directed interaction. The Yu prototype~\cite{zhuDesigningDigitalMindfulness2017} exemplifies this approach by avoiding explicit feedback, metrics, or instruction. Instead, it provides a minimal and open-ended interaction that invites sustained presence without evaluation. Unlike externalization-based systems, Yu does not construct an object of reflection. Instead, it removes conditions that reinforce identification with the self, such as performance monitoring or task completion.

Finally, inspired by buddhist rituals and mindfulness, \citet{mahDesigningRitualInteraction2020}'s artwork encourage participants to move away from an individualistic and self-focused orientation toward a more relational and compassionate awareness of other. Rather than emphasizing personal achievement or self-expression, the system encourages participants to cultivate generosity, loving-kindness, compassion, and ``other-centredness.''

Across these systems, different mechanisms for supporting change in self-perspective can be observed. Metaphorical externalization supports continuous reflection on internal states. Tangible interaction enables direct engagement with representations of thoughts and behaviors. Non-instrumental designs reduce self-referential processing by shifting attention toward ongoing experience. Despite these differences, all approaches converge on enabling a transition from identification with experience toward an observational, meta-aware stance.

\subsection*{Meta papers} We found 7 papers (10\%) that investigated, at a meta level, how tangible and embodied technologies can support mindfulness practice. Among these, three papers were review-oriented work~\cite{terzimehicReviewAmpAnalysis2019, markumMediatingSacredConfiguring2024, zhouTangibleAffectLiterature2024}. The remaining four papers consisted of research-through-design and theoretical contributions that explicitly addressed at least two or more mindfulness mechanisms proposed by \citet{holzelHowDoesMindfulness2011}. Following the recommendations of \citet{arkseyScopingStudiesMethodological2005}, we organize the findings thematically using mindfulness mechanisms as the analytic structure for synthesis.

\paragraph{Attention Regulation} We found four papers \cite{daudenroquetBodyMattersExploration2020, liCodesigningMagicMachines2023, liMeditationUnderstandingEveryday2024a, staabCanYouBe2024} collectively suggest that mindfulness technologies can be designed specifically to support this attentional loop without replacing the practitioner's own regulatory role. \citet{daudenroquetBodyMattersExploration2020} directly frames mind-wandering as a central challenge during meditation and proposes that technologies should support the transition back to mindful attention through haptic feedback and embodied metaphors that ``recreate mindful physical sensations during moments of mind-wandering''. For example, \citet{liCodesigningMagicMachines2023} propose \textit{Mindful Mug} that transforming the everyday act of drinking coffee into a mindful ritual through timed interruptions and focused engagement. Similarly, A \textit{Mindful Water Bottle} encourages users to pause and complete mindfulness activities before drinking water, thereby redirecting attention back to the present moment during routine daily activities. Likewise, \cite{liMeditationUnderstandingEveryday2024a} suggests that mindfulness technologies can function as attentional scaffolds not only during formal meditation but also throughout everyday activities. Participants described propose reminders, guided sessions, environmental structuring, and intentional interaction rituals as ways technologies can facilitate this process. Within \citet{holzelHowDoesMindfulness2011}'s framework, these mechanisms can be interpreted as supporting repeated attentional returning through ``micro-returns'' to awareness during ordinary routines. Finally, \cite{staabCanYouBe2024} contributes a situated perspective on attentional regulation by showing how devices can re-frame ordinary activities such as eating or listening to music into attentional anchors through staged prompts, reflective pauses, and body-focused cues. Although the paper does not strongly support the sustained concentration loop central to \citet{holzelHowDoesMindfulness2011}'s account of focused-attention meditation, it nevertheless demonstrates how contextual triggers and embodied prompts can interrupt automaticity and redirect attention back toward present-moment bodily sensation.

\paragraph{Body Awareness} We found three papers~\cite{daudenroquetBodyMattersExploration2020, liCodesigningMagicMachines2023, staabCanYouBe2024} broadly discuss tangible systems for body-awareness. \citet{daudenroquetBodyMattersExploration2020} argues that meditation involves distinct bodily sensations across stages and that grounding practices rely on increasing awareness of posture, breath, and internal sensation; accordingly, the authors propose that mindfulness technologies should not merely sense physiology passively but actively reconnect users with bodily experience through tactile and haptic actuation. \citet{liCodesigningMagicMachines2023} extended this notion by proposing magic machines: \textit{Magic Yoga Mat} and \textit{Rosary Bead Bracelet} that promotes tactile and sensory engagement through repetitive touch and physical manipulation, grounding attention in bodily sensation. Finally, \cite{staabCanYouBe2024} frame technology as a mediator for noticing, articulating, and interpreting bodily sensation phenomenologically. The paper demonstrates concrete approaches including body maps for locating sensations, multimodal expression through drawing, text, images, and stickers, reflective mindful eating activities, music-linked sensation exploration, and just-in-time interoceptive prompts. The paper further proposes combinations of sensing and expressive interaction, such as respiration-sensing wearables paired with tactile breathing rhythms, body maps generated from physiological sensing, or spatialized haptic feedback directing awareness toward regions of tension or calm.

\paragraph{Change in Perspective on the Self (Decentering)} 
Instead of helping users analyze what they are thinking, technologies could gently redirect attention toward the recognition that thinking itself is occurring. The notes specifically suggest subtle prompts or visualizations of mental activity as transient events, which aligns closely with \citet{holzelHowDoesMindfulness2011} description of mindfulness as fostering ``detachment from identification with the contents of consciousness'' and the development of an ``observer perspective.'' In \cite{liCodesigningMagicMachines2023} propose a magic machines:\textit{Magic Mirror}, a reflective object designed to display mindfulness prompts during everyday routines such as preparing in the morning or before bed. Because mirrors are objects that people encounter repeatedly every day, the system embeds moments of reflection directly into daily self-observation. Similarly, \cite{liMeditationUnderstandingEveryday2024a} extends this mechanism into everyday life by showing that mindfulness becomes integrated as an ongoing mode of awareness rather than functioning merely as a relaxation practice. Participants reportedly described increased awareness of thoughts, emotions, habits, and even technology use itself, alongside reduced automaticity in daily behavior. Across both papers, the shared implication for \citet{holzelHowDoesMindfulness2011} framework is that technologies supporting decentering should reduce identification with ongoing mental and behavioral processes while maintaining low conceptual load and help cultivating an observer perspective that operates continuously within everyday experience rather than only during formal meditation practice.

\begin{table}[t]
\centering
\caption{Classification of mindfulness-related device papers by device type and evaluation approach.}
\label{tab:mindfulness_device_eval_classification}

\begin{tabular}{p{2cm} p{3cm} p{8cm}}
\toprule
  Papers with tangible devices & Papers reporting on the evaluation of the device & \cite{MindPool2013}, \cite{vidyarthiInteractivelyMediatingExperiences2014}, \cite{sas2015meditaid}, \cite{Engagement_through_Embodiment_16}, \cite{macik2017breathinga}, \cite{paredes2017evaluating}, \cite{rooInnerGardenConnecting2017a}, \cite{seolDropBeatVirtual2017}, \cite{paredesJustBreatheIncar2018}, \cite{pryssPersonalizedSensorSupport2018}, \cite{semertzidisUnderstandingDesignPositive2019}, \cite{vianelloTANGAEONTangibleInteraction2019}, \cite{Mediscape2020}, \cite{aslanPiHeartsResonatingExperiences2020}, \cite{chinarevaLotusMediatingMindful2020}, \cite{choiAmbienBeatWristwornMobile2020a}, \cite{fooSoftRoboticCompression2020}, \cite{miriPIVPlacementPattern2020}, \cite{miriEvaluatingPersonalizableInconspicuous2020}, \cite{mahDesigningRitualInteraction2020}, \cite{daudenroquetInteroceptiveInteractionEmbodied2021a}, \cite{FirstPersonWalking2021}, \cite{sabinsonPlantHumanEmbodiedBiofeedback2021}, \cite{choiDesignEvaluationClippable2022}, \cite{cochraneBreathingScarfUsing2022}, \cite{EtherealPhenomena2022}, \cite{farrallManifestingBreathEmpirical2023a}, \cite{kuDisImmersionMindfulness2023}, \cite{tanMindfulMomentsExploring2023}, \cite{dublinJourneyInwardSomaesthetic2024}, \cite{ezerSomaestheticMeditationWearable2024a}, \cite{gemiciogluBreathePulsePeripheralGuided2024}, \cite{wang2024design}, \cite{AmbientPlantforMeditation2024}, \cite{ConsciousOrUnconsciousMeditation2025}, \cite{chenLivingBentoHeartbeatDriven2025}, \cite{choMindfulTouchMidair2025}, \cite{dublinWalkingMeditationMat2025a}, \cite{huangCoralMorphArtistic2025}, \cite{hyunVibroCushionDesignInclusive2025}, \cite{tanRunMeAdaptiveSound2025}, \cite{wangReflectingSoloDining2025}, \cite{BreathingInward2026}, \cite{EncouragingBreath2026}, \cite{Samten2026} \\ 
 & Papers not reporting on the evaluation of the device   & \cite{feijs2005design}, \cite{khotSWANDesigningCompanion2020}, \cite{KhongKhro2022}, \cite{EphemeralBreath2026}, \cite{SMArtBracelet2023} \\
 Papers with conceptual devices & devices developed through research through design(RtD) & \cite{vidyarthi2012sonic}, \cite{thiemeDesignPromoteMindfulness2013a}, \cite{aslanHoldMyHeart2016}, \cite{Sprite_Catcher_2017}, \cite{zhuDesigningDigitalMindfulness2017}, \cite{tenbhomerDesigningPersonalizedMovementbased2018}, \cite{huttonReMiNDImprovingEmotional2019}, \cite{daudenroquetBodyMattersExploration2020}, \cite{jungExploringAwarenessBreathing2021a}, \cite{liCodesigningMagicMachines2023}, \cite{liMeditationUnderstandingEveryday2024a}, \cite{staabCanYouBe2024} \\
 Review and survey papers & & \cite{terzimehicReviewAmpAnalysis2019}, \cite{markumMediatingSacredConfiguring2024}, \cite{zhouTangibleAffectLiterature2024} \\
\bottomrule
\end{tabular}
\end{table}

\subsection{\textbf{RQ2}: How are tangible and interactive systems evaluated for mindfulness?}

We found that 45 papers (68\%) reported empirical evaluations of the proposed devices, whereas 5 papers (7\%) described tangible systems without presenting any form of user or system evaluation which were works in progress. In addition to implemented artifacts, a subset of the literature focused on conceptual or speculative designs. Specifically, 12 papers (20\%) adopted research-through-design or co-design methodologies, contributing theoretical or design-oriented insights without evaluation. Furthermore, 3 papers (4\%) were identified as review articles, synthesising existing knowledge on tangible interaction rather than introducing new devices (see \autoref{tab:mindfulness_device_eval_classification} for the full list of papers). To address RQ2, we focus on the subset of papers that present and assess mindfulness-supporting devices and organise them according to their evaluation methodologies (aka, first row of \autoref{tab:mindfulness_device_eval_classification}). Through open coding of these methodologies, we structure the findings into three subthemes.


\begin{xltabular}{\columnwidth}{
p{0.2\columnwidth}
p{0.45\columnwidth}
p{0.3\columnwidth}
}
\caption{Psychological and affective evaluation measures used across studies}
\label{tab:psych_measures} \\
\toprule
\textbf{Construct} & \textbf{Questionnaire} & \textbf{Reported Paper} \\
\midrule
\endfirsthead

\multicolumn{3}{c}{\tablename\ \thetable{} -- continued from previous page} \\
\toprule
\textbf{Construct} & \textbf{Questionnaire} & \textbf{Reported Paper} \\
\midrule
\endhead

\midrule
\multicolumn{3}{r}{Continued on next page} \\
\endfoot

\bottomrule
\endlastfoot

Trait Mindfulness & Mindful Attention Awareness Scale (MAAS) 
    & \cite{tanRunMeAdaptiveSound2025} \\

    & Five Facet Mindfulness Questionnaire (FFMQ) 
    & \cite{rooInnerGardenConnecting2017a} (as a demographic) \\

State Mindfulness & Toronto Mindfulness Scale (TMS) 
    & \cite{rooInnerGardenConnecting2017a}, \cite{vianelloTANGAEONTangibleInteraction2019}, \cite{farrallManifestingBreathEmpirical2023a} \\

    & State Mindfulness Scale (SMS)
    & \cite{BreathingInward2026}\\

    & Mindfulness Attention Awareness Scale - State (MAAS-State)
    & \cite{daudenroquetInteroceptiveInteractionEmbodied2021a}, \cite{tanMindfulMomentsExploring2023}\\

    & Mindful Eating Questionnaire (MEQ)
    & \cite{chenLivingBentoHeartbeatDriven2025} \\

    & Embodied Mindfulness Questionnaire (EMQ)
    & \cite{EncouragingBreath2026} \\

Meditation Depth & Meditation Depth Questionnaire 
    & \cite{choMindfulTouchMidair2025} \\

Interoception & Multidimensional Assessment of Interoceptive Awareness (MAIA-2)
    & \cite{tanRunMeAdaptiveSound2025}, \cite{BreathingInward2026}, \cite{EncouragingBreath2026}  \\

    & Body Sensations Questionnaire 
    &  \cite{miriEvaluatingPersonalizableInconspicuous2020}\\

State Anxiety & State-Trait Anxiety Inventory (STAI-1968) 
    & \cite{fooSoftRoboticCompression2020}, \cite{seolDropBeatVirtual2017} \\

    & STAI-X-1 
    & \cite{choiDesignEvaluationClippable2022} \\

    & STAI-YA 
    & \cite{rooInnerGardenConnecting2017a} \\

    &  Short Six-Item STAI (1992)
    & \cite{miriEvaluatingPersonalizableInconspicuous2020}, \cite{gemiciogluBreathePulsePeripheralGuided2024}\\

    & Short Six-Item STAI (2009) 
    & \cite{farrallManifestingBreathEmpirical2023a} \\

Anxiety & The Anxiety Sensitivity Index 
    &  \cite{miriEvaluatingPersonalizableInconspicuous2020}\\

    & Beck Anxiety Inventory (BAI)
    & \cite{seolDropBeatVirtual2017} \\

    & Hamilton Anxiety Rating Scale (HAM-A)
    & \cite{seolDropBeatVirtual2017} \\

    & Depression Anxiety Stress Scale (DASS)
    & \citet{EncouragingBreath2026} \\

Stress & Perceived Stress Scale (PSS) 
    & \cite{choiDesignEvaluationClippable2022}, \cite{Mediscape2020} \\

    & Perceived Stress (Custom Scales)
    & \cite{paredes2017evaluating}, \cite{paredesJustBreatheIncar2018}, \cite{choiAmbienBeatWristwornMobile2020a}\\

Affect & Positive and Negative Affect Schedule (PANAS) 
    & \cite{semertzidisUnderstandingDesignPositive2019}, \cite{gemiciogluBreathePulsePeripheralGuided2024} \\

    & Positive and Negative Affect Schedule Extended (PANAS-X) 
    & \cite{semertzidisUnderstandingDesignPositive2019} \\

    & Trait Meta-Mood Scale (TMMS-24)
    & \cite{huangCoralMorphArtistic2025}\\

Affective State & Visual Analogue Scale (VAS) 
    & \cite{farrallManifestingBreathEmpirical2023a} \\

Affective Response & Self-Assessment Manikin (SAM) 
    & \cite{vianelloTANGAEONTangibleInteraction2019}, \cite{choMindfulTouchMidair2025}, \cite{huangCoralMorphArtistic2025} \\

    & Unipolar Valence Model
    &  \cite{miriEvaluatingPersonalizableInconspicuous2020}\\

    & Brief Mood Introsepction Scale (BMIS)
    & \cite{huangCoralMorphArtistic2025} \\

Mood States & Profile of Mood States (POMS) 
    & \cite{choMindfulTouchMidair2025} \\

    & Profile of Mood States Short Form (POMS-SF)
    & \cite{cochraneBreathingScarfUsing2022} \\

Arousal & Pre-Sleep Arousal Scale
    & \cite{semertzidisUnderstandingDesignPositive2019} \\

Emotion Regulation & Emotion Regulation Questionnaire
    &  \cite{miriEvaluatingPersonalizableInconspicuous2020}\\

    & Difficulties in Emotion Regulation Questionnaire
    &  \cite{miriEvaluatingPersonalizableInconspicuous2020}\\

Cognitive Load & NASA Task Load Index (NASA-TLX) 
    &  \cite{choiDesignEvaluationClippable2022}, \cite{gemiciogluBreathePulsePeripheralGuided2024} \\

Flow & Flow State Scale (FSS) 
    & \cite{fooSoftRoboticCompression2020}, \cite{EncouragingBreath2026} \\

    & Flow State Questionnaire (PPL-FSQ) 
    & \cite{farrallManifestingBreathEmpirical2023a} \\

Motivation & Intrinsic Motivation Inventory (IMI) 
    & \cite{farrallManifestingBreathEmpirical2023a}, \cite{tanRunMeAdaptiveSound2025} \\

    & Situational Motivation Scale (SIMS)
    & \cite{EncouragingBreath2026} \\

Well-being & Short Warwick-Edinburgh Mental Wellbeing Scale (SWEMWBS) 
    & \cite{farrallManifestingBreathEmpirical2023a} \\

    & Warwick-Edinburgh Mental Wellbeing Scale (WEMWBS) 
    & \cite{EncouragingBreath2026} \\

    & World Health Organization-Five Well-Being Index (WHO-5)
    & \cite{Mediscape2020} \\

Personality Traits & Big Five Inventory (BFI) 
    &   \cite{miriEvaluatingPersonalizableInconspicuous2020}, \cite{choiDesignEvaluationClippable2022} \\

Multi-scale & Game Experience Questionnaire*
    & \cite{aslanPiHeartsResonatingExperiences2020} \\

Attention & Measure of Attention Focus (MAF)
    & \cite{tanRunMeAdaptiveSound2025} \\

Custom Scale & Custom
    & \cite{sas2015meditaid}, \cite{Engagement_through_Embodiment_16}, \cite{paredes2017evaluating}, \cite{macik2017breathinga}, \cite{paredesJustBreatheIncar2018}, \cite{Mediscape2020}, \cite{chinarevaLotusMediatingMindful2020}, \cite{choiAmbienBeatWristwornMobile2020a}, \cite{choiDesignEvaluationClippable2022}, \cite{tanMindfulMomentsExploring2023}, \cite{gemiciogluBreathePulsePeripheralGuided2024}, \cite{chenLivingBentoHeartbeatDriven2025}, \cite{miriPIVPlacementPattern2020}, \cite{sabinsonPlantHumanEmbodiedBiofeedback2021}, \cite{mahDesigningRitualInteraction2020}, \cite{EtherealPhenomena2022}, \cite{ConsciousOrUnconsciousMeditation2025}, \cite{EncouragingBreath2026} \\

\end{xltabular}

\subsection*{Psychological, Cognitive and Behavioral Evaluation}

Surprisingly, we found only 9 papers~\cite{tanRunMeAdaptiveSound2025, rooInnerGardenConnecting2017a, vianelloTANGAEONTangibleInteraction2019, farrallManifestingBreathEmpirical2023a, daudenroquetInteroceptiveInteractionEmbodied2021a, tanMindfulMomentsExploring2023, chenLivingBentoHeartbeatDriven2025, BreathingInward2026, EncouragingBreath2026} that employed validated questionnaires to assess mindfulness. Trait mindfulness was reported in 3~papers, using the Five Facet Mindfulness Questionnaire (FFMQ)~\cite{baer2006using}, Mindful Attention Awareness Scale (MAAS)~\cite{brown2003benefits}. 
State mindfulness was assessed in 8 papers, most commonly using the Toronto Mindfulness Scale (TMS)~\cite{lau2006toronto}, and less frequently the State Mindful Attention Awareness Scale-State(MAAS-State)~\cite{MAAS_State}, Mindful Eating Questionnaire (MEQ)~\cite{MEQ_Development} and Embodied Mindfulness Questionnaire~\cite{khoury2023embodied}

Beyond mindfulness, a wide range of psychological constructs are evaluated. Anxiety was the most frequently assessed construct, primarily measured using variants of the State-Trait Anxiety Inventory (STAI)~\cite{STAI_x_form}, including short-form adaptations~\cite{marteau1992development, tluczek2009support}. \citet{seolDropBeatVirtual2017} did not specify which version of the STAI was employed; therefore, it was grouped under the primary STAI category. In addition, \citet{seolDropBeatVirtual2017} reported anxiety using the Hamilton Anxiety Rating Scale (HAM-A)~\cite{thompson2015hamilton} and the Brief Symptom Inventory (BSI) anxiety scale~\cite{fydrich1992reliability}. Stress and broader negative affect are also commonly captured, using the Depression Anxiety Stress Scale (DASS-21)~\cite{beaufort2017depression} and the Perceived Stress Scale (PSS)~\cite{roberti2006further}. Positive and negative affect are reported using the Positive and Negative Affect Schedule (PANAS)~\cite{watson1988development}. Also in a single case \citet{seolDropBeatVirtual2017} reported depression using the Montgomery–Asberg Depression Rating Scale (MADRS)~\cite{montgomery1979new}. General well-being and motivation-related constructs are less frequently assessed. \citet{farrallManifestingBreathEmpirical2023a} evaluated mental well-being using the Short Warwick-Edinburgh Mental Wellbeing Scale (SWEMWBS)~\cite{maheswaran2012evaluating}, while motivation and engagement using the Intrinsic Motivation Inventory (IMI)~\cite{gonzalez2020psychological}. Personality traits are considered through the Big Five Inventory (BFI)~\cite{john1991big}.

Several studies also relied on more immediate or subjective affective measures using Visual Analogue Scales (VAS)~\cite{ainsworth2017testing} to capture transient affective states, and the Self-Assessment Manikin (SAM)~\cite{bradley1994measuring} for dimensional emotion assessment. Mood states are assessed in using the Abbreviated Profile of Mood States (POMS)~\cite{grove1992preliminary} and Short form of Profile of Mood States (POMS-SF)~\cite{curran1995short}.

Also, experiential and task-related constructs are reported. Flow was measured in 3~papers using the Flow State Scale (FSS)~\cite{jackson1996FSS} or the Flow State Questionnaire (PPL-FSQ)~\cite{magyarodi2013pplfsq}, while cognitive load and workload are assessed in 2~papers using the NASA Task Load Index (NASA-TLX)~\cite{hart1988development}.

Finally, a substantial proportion of the reviewed papers relied on adapted self-report measures. These custom scales typically capture embodied, experiential, and interaction-specific aspects of engaging with tangible mindfulness devices, particularly dimensions that existing mindfulness questionnaires do not adequately operationalize. For example, \citet{sas2015meditaid} argued that no validated questionnaire adequately captured the self-regulation of attention during meditation, and therefore introduced three custom items assessing participants' perceived stillness, frequency of attentional drift, and percentage of time their mind remained still during meditation. Similarly, \citet{Engagement_through_Embodiment_16} evaluated mindful engagement using Rozendaal's Richness, Control \& Engagement (RC\&E) framework~\cite{rozendaal2007designing}, emphasizing experiential qualities of interaction rather than mindfulness outcomes alone. Several studies focusing on breathing-based or biofeedback-driven systems also developed tailored measures to examine how participants perceived and interacted with the device itself. For instance, \citet{choiDesignEvaluationClippable2022} designed Likert-scale items assessing perceived tactile feedback, awareness of breathing, effort involved in breathing regulation, and perceived influence of the device on breathing control. Likewise, \citet{gemiciogluBreathePulsePeripheralGuided2024} measured participants' attention to, and adherence with, airflow-guided breathing cues during a concurrent task.

Custom measures were also commonly used to capture immediate experiential or situational states during device use. In the context of in-car mindfulness interventions, \citet{paredes2017evaluating} repeatedly prompted participants to report their current ``stress'' and ``concentration levels'' during driving, while in their later work~\cite{paredesJustBreatheIncar2018} authors extended this approach by additionally assessing the attentional demands imposed by the guidance system. In contrast, some studies reported adapting existing validated scales to better fit the tangible interaction context rather than introducing entirely new instruments. For example, \citet{chenLivingBentoHeartbeatDriven2025} explicitly stated that established questionnaires were modified to align with the aims of the study and the characteristics of the device-mediated experience.




\begin{table}[t]
\caption{Physiological and behavioral evaluation measures used across studies}
\label{tab:phys_measures}
\centering
\begin{tabularx}{\columnwidth}{l X X}
\toprule
\textbf{Construct} & \textbf{Measure} & \textbf{Reported Paper} \\
\midrule

Respiration & Breathing Rate 
& \cite{macik2017breathinga}, \cite{paredesJustBreatheIncar2018}, \cite{fooSoftRoboticCompression2020}, \cite{miriPIVPlacementPattern2020}, \cite{choiDesignEvaluationClippable2022}, \cite{gemiciogluBreathePulsePeripheralGuided2024}, \cite{wang2024design} \\

& Respiratory Wave Amplitude 
& \cite{choiDesignEvaluationClippable2022}, \cite{gemiciogluBreathePulsePeripheralGuided2024} \\

& Percentage of Time Slow Breathing 
& \cite{gemiciogluBreathePulsePeripheralGuided2024}\\

Cardiovascular Activity & Heart Rate (HR/ECG)
& \cite{paredesJustBreatheIncar2018}, \cite{chinarevaLotusMediatingMindful2020}, \cite{choiAmbienBeatWristwornMobile2020a}, \cite{fooSoftRoboticCompression2020}, \cite{miriPIVPlacementPattern2020}, \cite{wang2024design}, \cite{BreathingInward2026} \\

& Heart Rate Variability (HRV) 
& \cite{wang2024design} \\

& RMSSD 
& \cite{paredesJustBreatheIncar2018}, \cite{farrallManifestingBreathEmpirical2023a}, \cite{gemiciogluBreathePulsePeripheralGuided2024}, \cite{BreathingInward2026}, \cite{EncouragingBreath2026} \\

& SDNN 
& \cite{choiAmbienBeatWristwornMobile2020a} \\

Electrodermal Activity & Electrodermal Activity (EDA) 
& \cite{paredesJustBreatheIncar2018}, \cite{fooSoftRoboticCompression2020}, \cite{miriPIVPlacementPattern2020}, \cite{choiDesignEvaluationClippable2022}, \cite{farrallManifestingBreathEmpirical2023a}, \cite{tanMindfulMomentsExploring2023} \\

EEG & Frequency bands
& \cite{sas2015meditaid}, \cite{semertzidisUnderstandingDesignPositive2019} \\

Body Temperature & Temperature
& \cite{miriPIVPlacementPattern2020} \\

Attention & Sustained Attention to Response Task (SART)
& \cite{tanMindfulMomentsExploring2023} \\

Walking Speed & Custom
& \cite{pryssPersonalizedSensorSupport2018}\\

Body maps & Body maps
& \cite{cochraneBreathingScarfUsing2022}\\

\bottomrule
\end{tabularx}
\end{table}

\subsection*{Physiological and Behavioral Evaluation}

Across the reviewed studies (see \autoref{tab:phys_measures}), a range of physiological and behavioral measures are used to assess user state. Respiratory measures are the commonly captured across papers through breathing rate, typically expressed as breaths per minute (BPM), and respiratory wave amplitude, which reflects the depth of breathing. Some studies additionally reported derived measures such as the proportion of time spent in slow, controlled breathing.

Cardiovascular activity was most most commonly captured through heart rate (HR) and heart rate variability (HRV). Heart rate (HR) is typically derived from electrocardiography (ECG) signals and reported in beats per minute (bpm), reflecting overall cardiac activity. In contrast HRV captures variations in the time interval between successive heartbeats and reported in two main ways: root mean square of successive differences (RMSSD) and standard deviation of NN intervals (SDNN). Most papers that reported HRV used RMSSD value, which reflects short-term parasympathetic activity. However, we found difference in reporting HRV values. For example, both \cite{paredesJustBreatheIncar2018} and \cite{farrallManifestingBreathEmpirical2023a} used the time-domain HRV metric RMSSD derived from ECG recordings to assess autonomic and emotional regulation. First \citet{paredesJustBreatheIncar2018} treated RMSSD a direct physiological outcome measure and baseline-normalized across participants. In contrast, rather than analysing absolute RMSSD values alone, \citet{farrallManifestingBreathEmpirical2023a} reported baseline-corrected changes in RMSSD (\(\Delta RMSSD\)) over time, using moving-average smoothing and regression modelling to evaluate group-level intervention effects. In contrast to RMSSD, \citet{leeAmbientBreathUnobtrusiveJustintime2021} reported SDNN, a measure associated with physiological resilience to stress using time-domain analysis of heart rate and R–R interval variation data, where mean normal-to-normal (NN) intervals and variance between NN intervals were calculated to derive the SDNN.

Also we found Electro-dermal activity (EDA) as another frequently used measure to capture changes in skin conductance associated with sympathetic nervous system activation and emotional arousal. In EDA signals, tonic activity refers to the slower-changing baseline level of skin conductance, while phasic activity refers to faster transient responses associated with momentary emotional or physiological reactions. We found differences in how they quantified and analyzed these components (discrete and continuous analysis). For example, \citet{paredesJustBreatheIncar2018} used a feature-based approach, focusing on the average tonic skin conductance level and the number of phasic peaks detected during the experiment by treating stress responses as discrete events that can be counted and summarized. In contrast, \citet{miriEvaluatingPersonalizableInconspicuous2020} argued that conventional peak detection can be inaccurate when multiple skin conductance responses overlap in time. Instead, they applied Continuous Decomposition Analysis (CDA), which models the phasic activity continuously over time and calculates the average phasic driver activity within fixed response windows.

Neural activity was assessed using electroencephalography (EEG), with analyses focusing on frequency bands (e.g., alpha, beta) to infer cognitive and affective states. Body temperature was occasionally used as an indicator of physiological regulation. Behavioral and cognitive measures are also present, including attention assessed through the Sustained Attention to Response Task (SART), and physical activity indexed via walking speed using custom metrics.

\begin{table}[!t]
\caption{Usability and user interaction quality measures used across studies}
\label{tab:ux_measures}
\centering
\begin{tabularx}{\columnwidth}{l X X}
\toprule
\textbf{Construct} & \textbf{Measure} & \textbf{Reported Paper} \\
\midrule

Overall Usability 
& System Usability Scale (SUS) 
& \cite{rooInnerGardenConnecting2017a} \\

User Experience (Custom) 
& Custom Questions (SEQ) 
& \cite{paredes2017evaluating}, \cite{rooInnerGardenConnecting2017a}, \cite{paredesJustBreatheIncar2018}, \cite{vianelloTANGAEONTangibleInteraction2019}, \cite{chinarevaLotusMediatingMindful2020}, \cite{gemiciogluBreathePulsePeripheralGuided2024}, \cite{wang2024design}, \cite{sabinsonPlantHumanEmbodiedBiofeedback2021}\\

Comparative Preference 
& Device Preference Comparison 
& \cite{vianelloTANGAEONTangibleInteraction2019}, \cite{choiAmbienBeatWristwornMobile2020a} \\

Experienced Embodiment & Interaction Vocabulary
& \cite{Engagement_through_Embodiment_16} \\

\bottomrule
\end{tabularx}
\end{table}

\subsection*{Usability and User Interaction Quality}

Across the reviewed studies, usability and user interaction quality are assessed using a combination of standardized instruments and custom-designed measures. For example, \citet{rooInnerGardenConnecting2017a} reported overall usability using the System Usability Scale (SUS)~\cite{bangor2008empirical, lewis2018system}, a widely adopted questionnaire that provides a global measure of perceived system usability based on user-reported ease of use, efficiency, and satisfaction~\cite{lewis2018system}. Also 8 papers user experience custom questions, often implemented as single-item or short-form assessments such as the Single Ease Question (SEQ). These measures typically focus on users' immediate perceptions of interaction quality, including ease, comfort, and overall experience during task execution. Also 2 papers \cite{vianelloTANGAEONTangibleInteraction2019, choiAmbienBeatWristwornMobile2020a} employed comparative preference measures, where participants directly evaluated and expressed preferences between multiple devices or interaction modalities.

\begin{table}[!t]
\caption{Subjective experience and perceived impact methods used across studies}
\label{tab:subjective_measures}
\centering
\begin{tabularx}{\columnwidth}{l X X}
\toprule
\textbf{Construct} & \textbf{Method} & \textbf{Reported Paper} \\
\midrule

User Experience (Qualitative) & Interviews 
& \cite{vidyarthi2012sonic}, \cite{MindPool2013}, \cite{vidyarthiInteractivelyMediatingExperiences2014}, \cite{sas2015meditaid}, \cite{macik2017breathinga}, \cite{paredes2017evaluating}, \cite{rooInnerGardenConnecting2017a}, \cite{paredesJustBreatheIncar2018}, \cite{semertzidisUnderstandingDesignPositive2019}, \cite{vianelloTANGAEONTangibleInteraction2019}, \cite{aslanPiHeartsResonatingExperiences2020}, \cite{chinarevaLotusMediatingMindful2020}, \cite{choiAmbienBeatWristwornMobile2020a}, \cite{fooSoftRoboticCompression2020}, \cite{mahDesigningRitualInteraction2020}, \cite{Mediscape2020}, \cite{FirstPersonWalking2021}, \cite{daudenroquetInteroceptiveInteractionEmbodied2021a}, \cite{farrallManifestingBreathEmpirical2023a}, \cite{ezerSomaestheticMeditationWearable2024a}, \cite{dublinJourneyInwardSomaesthetic2024}, \cite{gemiciogluBreathePulsePeripheralGuided2024}, \cite{wang2024design}, \cite{chenLivingBentoHeartbeatDriven2025}, \cite{dublinWalkingMeditationMat2025a}, \cite{hyunVibroCushionDesignInclusive2025}, \cite{tanRunMeAdaptiveSound2025}, \cite{wangReflectingSoloDining2025}, \cite{sabinsonPlantHumanEmbodiedBiofeedback2021}, \cite{EtherealPhenomena2022}, \cite{AmbientPlantforMeditation2024}, \cite{BreathingInward2026}, \cite{Samten2026} \\

& Micro-phenomenological Interviews 
& \cite{kuDisImmersionMindfulness2023}, \cite{choMindfulTouchMidair2025} \\

& Diary-study
& \cite{macik2017breathinga} \\

Expert / Practitioner Perspective & Interviews with clinicians and/or mindfulness practitioners 
& \cite{macik2017breathinga}, \cite{thiemeDesignPromoteMindfulness2013a} \\

\bottomrule
\end{tabularx}
\end{table}

\subsection*{Subjective Experience and Perceived Impact}

Across the reviewed studies, subjective experience and perceived impact are primarily explored through qualitative methods that capture first-person accounts of interaction. The most common approach involved semi-structured interviews reported in 33 papers. The semi-structured interviews had first to use the system and then to reflect on their experiences, perceptions, and emotional responses to the system in depth. In contrast,  two studies \cite{kuDisImmersionMindfulness2023, choMindfulTouchMidair2025} specifically mentioned they employed micro-phenomenological interviews, a more structured qualitative technique designed to elicit detailed descriptions of lived experience. \citet{ezerSomaestheticMeditationWearable2024a} reported conducting pilot implementations of both micro-phenomenological and standard semi-structured interviews and ultimately adopted the standard semi-structured format with questions adopted from the the State Mindfulness Scale (SMS)~\cite{tanayStateMindfulnessScale2013}.

To measure longitudinal fitting to everyday life, \citet{macik2017breathinga} carried out a fourteen-day diary study to investigate how the ``Breathing Friend'' device could support stress reduction in everyday life. Four participants used the portable breathing artifact in natural settings, while recording their experiences in daily diaries and post-study interviews.

In addition to end-user perspectives, a \citet{macik2017breathinga} conducted semi-structured interviews with six therapists to evaluate the potential therapeutic effect of the device with psychologists, medical professionals, and yoga instructors, viewed the device positively and described it as calming and helpful for improving awareness of mindful breathing. In contrast with this \cite{thiemeDesignPromoteMindfulness2013a} used a proxy interviews and collaborative design approach, where researchers worked closely with therapists, nurses, clinical managers, and hospital staff to design the ``Spheres of Wellbeing'' for vulnerable women in secure psychiatric care. The authors report that the proxy approach was necessary as direct access to patients was ethically restricted due to the sensitivity of the user group.

\begin{figure}
    \centering
    \includegraphics[width=1\linewidth]{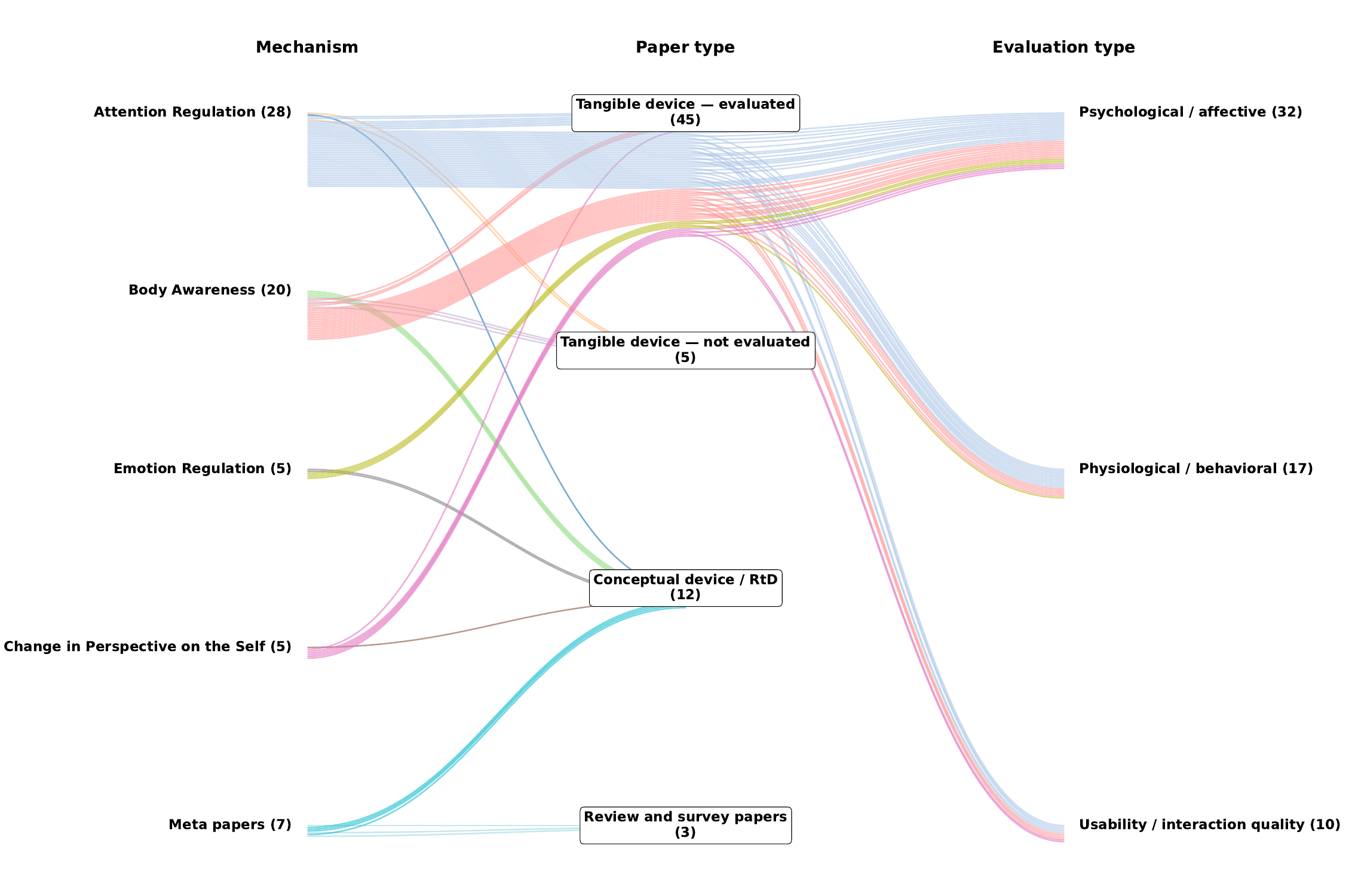}
    \caption{Sankey Diagram showing the progression and distribution of the papers between the mechanism of action of mindfulness and evaluation methods used}
    \Description{Sankey Diagram showing the progression and distribution of the papers between the mechanism of action of mindfulness and evaluation methods used}
    \label{fig:sunkey_diagram}
\end{figure}

\section{Consultation with Stakeholders}
\label{sec:consultation}
To validate the scoping review, we conducted an in-person consultation activity involving key stakeholders, including mindfulness experts, recent participants in the Mindfulness-Based Stress Reduction (MBSR) program, and HCI practitioners. In this scoping review, we prioritized RQ1 (i.e., design space), as the investigation of RQ2 (i.e., evaluation methods) was intended to build upon the insights derived from RQ1. To maintain methodological rigor, we treated the consultation as a focus group activity as methodological guidance for conducting such consultations remains limited~\cite{levac2010scoping}. In light of \citet{buus2022arksey}, who identify the omission of participant details, ethics, study design, and analysis procedures (including coder positionality) as a common limitation of consultation activities, we explicitly report these elements. However, to preserve the flow of the main text, detailed methodological discussions are provided in the \hyperref[sec:Appendix_3]{Appendix IV}, while only the findings are presented here.

We also recognized that directly presenting the scoping review findings to stakeholders could unintentionally constrain ideation by anchoring discussions to existing literature. This concern was particularly relevant as many prior studies are themselves products of research-through-design (RtD) processes and therefore already embody established design assumptions and solution spaces. To avoid limiting participants' ability to think beyond the literature, stakeholders first engaged in an ideation activity, where they were encouraged to generate a wide range of ideas, including unconventional concepts, before being exposed to findings from the review. We subsequently compared these stakeholder-generated ideas with the literature to identify areas of overlap and gaps. Accordingly, the following section should be understood as an examination of the robustness and consistency of the existing literature through stakeholder-generated perspectives, rather than as an attempt to claim novelty.

\begin{table*}
\centering
\begin{tabular}{lccp{9cm}}
\toprule
\textbf{Person} & \textbf{Age} & \textbf{Gender} & \textbf{Background \& Experience} \\ 
\midrule
G1\_H1 & 23 & Female & HCI Practitioner with experience in prototyping, 3D printing, PCB design and 1yr of experience in Yoga \\ 
G1\_M1 & 28 & Female & MBSR participant with 3 years since MBSR and MBCT 3 months ago, carriers out research on contemplative sciences \\ 
G1\_M2 & 21 & Female & MBSR participant with 2 years since MBSR, regular practitioner \\ 
G2\_M1 & 56 & Male & Mindfulness Expert with 30 years experience; runs a mindfulness company and organizes Buddhist meditation retreats; certified MBSR facilitator \\ 
G2\_H1 & 24 & Female & HCI Practitioner with experience in HCI and over 4 years of meditation practice \\ 
G3\_H1 & 21 & Female & HCI Practitioner with background in prototyping and PCB design \\ 
G3\_M1 & 49 & Male & Mindfulness Expert with 25 years experience; runs a mindfulness meditation company for workplaces \\ 
G3\_M2 & 34 & Female & MBSR participant with 4 years since MBSR \\ 
\bottomrule
\end{tabular}
\caption{Overview of focus group participants.}
\label{tab:participants}
\end{table*}

\subsection{Findings}

\begin{figure*}
    \centering
    \includegraphics[width=1\linewidth]{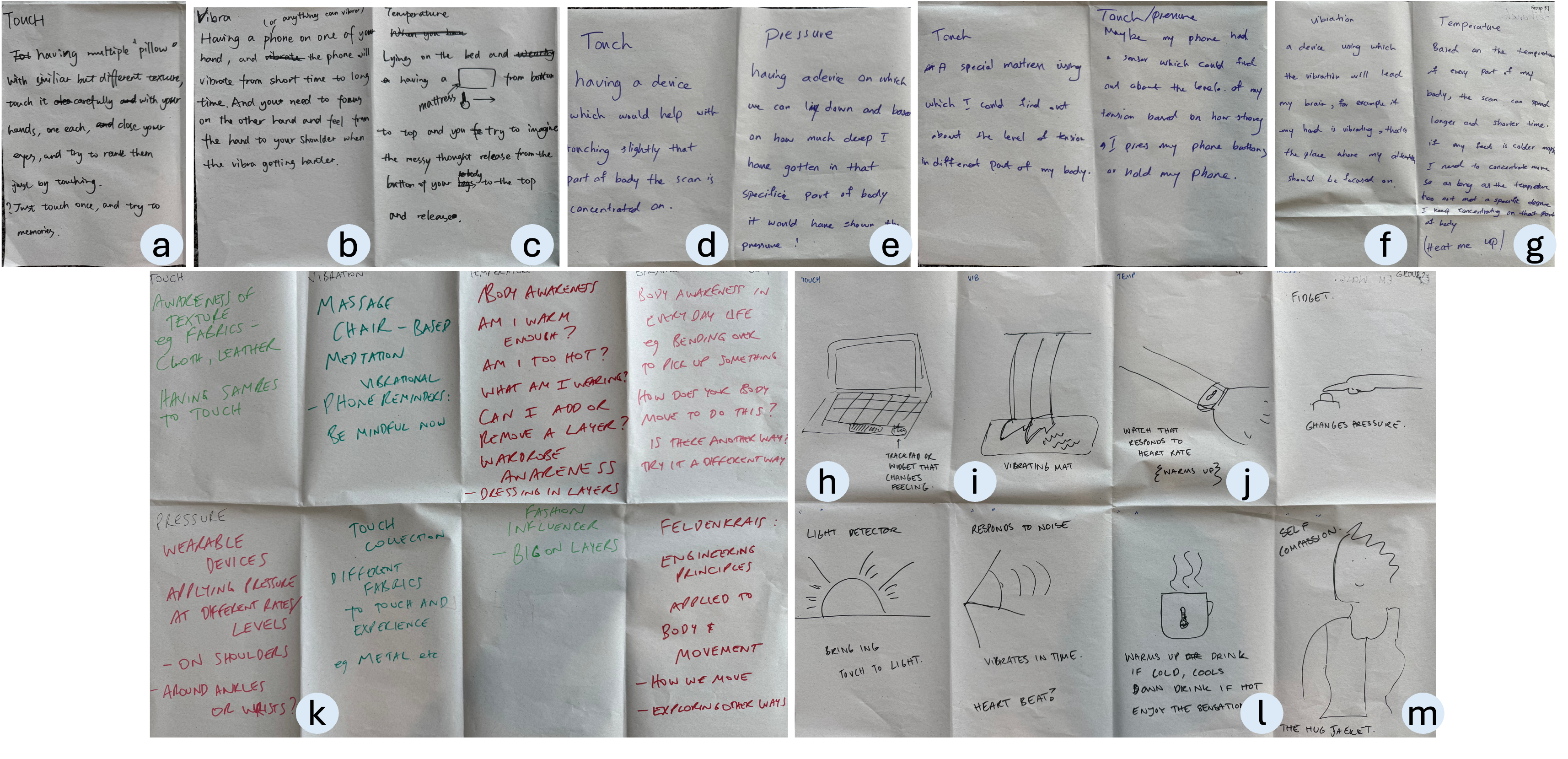}
    \caption{Mindfulness devices proposed by participants during the user study. (a) A pillow with different texture that can elicit different emotions (b) The interval vibrating device gradually spaces reminders to build independent focus, while the temperature-shifting mattress (c) uses rising warmth as a metaphor for releasing stress. The body-scan pressure device (d) and tension-mapping mattress (e) guide awareness to specific bodily regions, encouraging relaxation and interoceptive awareness. The stress-detecting pressure button (f) reflects emotional states through touch intensity, while the texture-changing object (g) fosters curiosity and adaptability by shifting tactile sensations. Gentle reminders from the vibrating leg mat (h) support mindfulness in daily routines, and the heart-responsive watch (i) connects physiological states with timely cues for self-regulation. Finally, the temperature-adaptive drink holder (j) turns an ordinary sip into a mindful ritual, grounding attention through comfort and sensory appreciation.}
    \label{fig:design_space}
\end{figure*}

In this section, we present the findings from the focus group extending RQ1 (i.e. design space), organized into three holistic categories.  The first concerns {what} participants view of mindfulness and challenges. Second, concerns \textit{what} participants expected in terms of the feedback provided by a tangible device. The third focuses on \textit{how} participants anticipated the device to generate feedback.

\begin{figure}[t]
    \centering
    \includegraphics[width=\linewidth]{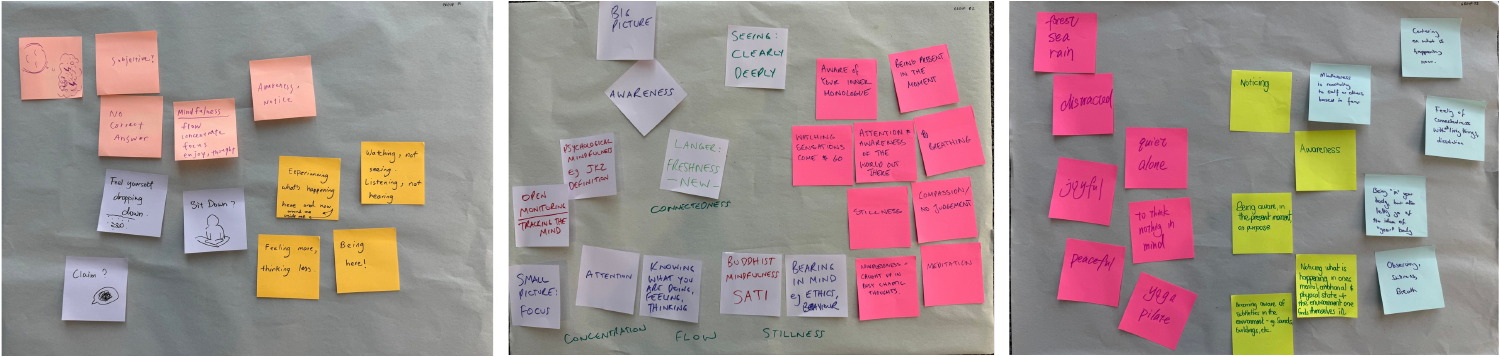}
    \caption{The participants reported diverse perspectives for definitions of mindfulness, ranging from cognitive and psychological definitions like ``awareness,'' ``attention,'' and ``stillness,'' to experiential and emotional interpretations such as ``being here,'' ``joyful,'' ``peaceful,'' and ``awareness of body and breath.''}
    \label{fig:appendix_definitions}
\end{figure}

\begin{figure}[t]
    \centering
    \includegraphics[width=\linewidth]{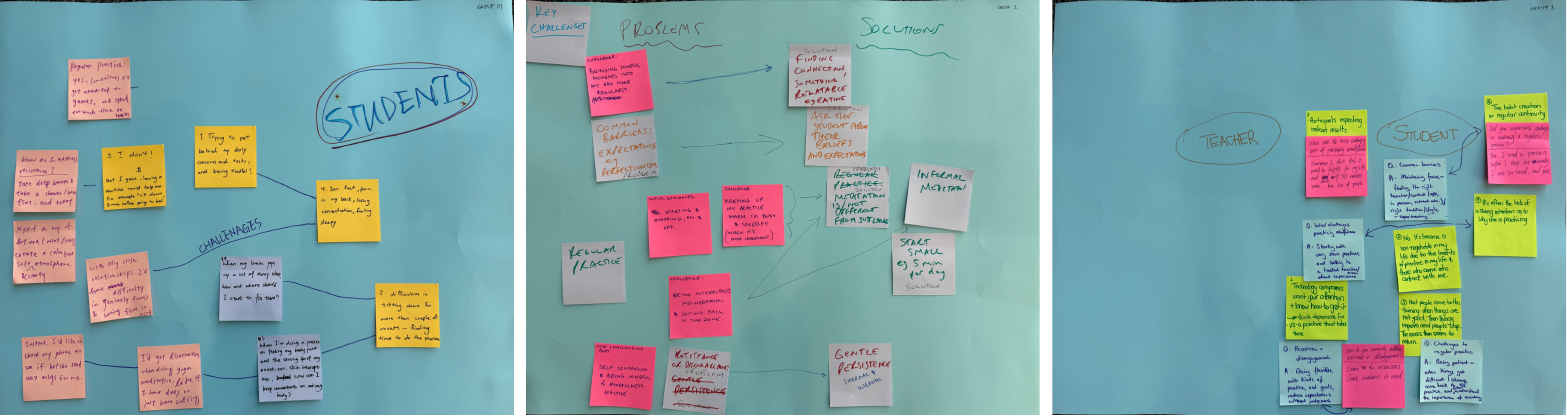}
    \caption{The participants highlighted student-related challenges such as lack of time, overthinking, low motivation. The ``Problems'' included distractions and unrealistic expectations and the ``Solutions'' included informal meditation, self-compassion, and gentle persistence}
    \label{fig:appendix_challenges}
\end{figure}

First, participants expressed diverse perspectives on the meaning of mindfulness, describing it through cognitive and psychological concepts such as awareness, attention, and stillness, as well as experiential and emotional interpretations including being present, joy, peace, and awareness of body and breath (see \autoref{fig:appendix_definitions}). Participants also highlighted several student-related challenges to mindfulness practice, including lack of time, overthinking, and low motivation, while identifying common problems such as distractions and unrealistic expectations and suggesting strategies such as informal meditation, self-compassion, and gentle persistence as potential solutions (see \autoref{fig:appendix_challenges}).

\paragraph{Mindfulness devices as adaptive guides for sustaining practice}

Second, participants acknowledged that novice meditators experience difficulty maintaining a meditation practice for three main reasons. First, they often hold unrealistic expectations regarding outcomes. For example, one participant contrasted mindfulness with more visibly measurable goals, stating, ``\textit{Losing weight is better, you can see your body shape changes. But for the mindfulness, you cannot tell whether you are getting better or not} (G1\_H1)''. This highlights the challenge of perceiving tangible progress in mindfulness practice.

Second, participants described difficulties in sustaining mindfulness practice due to distractions in everyday life. As one participant explained, ``\textit{I felt like my initial challenges faced when I did the MBSR course when we started really long practices like half-hour practices. I have found it really hard even if I had quite a bit of experience in meditation before with what is going on in my life. Everyone one in the MBSR course found it challenging and it would have been better if we had option for a shorter practices in the beginning and also like a common barrier maintaining the focus} (G3\_M2)'' suggesting that maintaining focus during practice can be particularly challenging for novices.

Third, novices may misunderstand the nature of mindfulness practice and the time required to experience meaningful benefits. As explained by G3\_M1:

\begin{quote}
    You know something that is well known about mindfulness, which is not correct, is that you don't feel calm when doing mindfulness. You will get back to your body, how you feel, how you hear, how you see. So, it's more about getting back to your body. And it doesn't necessarily bring calmness or pleasant feelings [...] but maybe in long term, people experience calmness by doing mindfulness, but I guess in the first years, it's just practice.
\end{quote}

To address these challenges, participants acknowledged that the device should be designed to support meditators’ \emph{needs, challenges,} and \emph{level of expertise}. In particular, participants emphasized that novice meditators require a \emph{structured approach} to mindfulness practice. This need arises from above three primary factors: lack of understanding, the desire for immediate results and the lack of motivation resulting from difficulties in perceiving progress.

As an approach, participants proposed two complementary forms of structure. (1) Within a session, devices could anchor attention, such as applying vibration or temperature during body scanning to guide focus. (2) Across sessions, devices could act as reminders, helping users build routines: ``It's like a reminder [to meditate]'' (G2\_M1). Such scaffolds were seen as reducing reliance on internal motivation and supporting the formation of habits.

Regarding the desire for immediate results, designers (of devices) face the challenge of tracking, as it compromises the non-striving nature of mindfulness. If tracking is necessary, participants suggested two strategies to address this issue. One approach is to avoid explicitly disclosing goals by not displaying performance scores, as this can create performance anxiety. Instead, they recommended offering positive encouragement in real-time based on the meditator's performance. Second, to word physiological feedback, like heart rate or HRV, which participants described as a helpful reassurance of one's state and even an external validator that the practice was ``working.''

\begin{quote}
   \textit{ It [our device] makes us judgmental about our performance because it will ask us to reach some degrees, some temperatures in our body. So in that way, it might put some pressure on us... So in order to take over that problem,[...] instead of giving the participants the temperature of their body, we give constant positive feedbacks about how they're doing. [...] if I'm doing better, if I'm making progress, I will be given positive feedback about what I've done instead of just telling me you need to reach that degree.}~(G1\_M1)
\end{quote}

In contrast, the strongly suggested approach was to completely remove any progress tracking and create devices with the idea of bringing into focus just the mindful moments.

\begin{quote}
\textit {I think one of the things that stood out for me was just trying to stay away from tracking because I think that kind of leads to anxiety [...] but rather bringing into focus just the simple joys of moments }(G3\_M2)
\end{quote}

Participants also proposed tailoring the device's objective according to the meditator's level of expertise. For novice practitioners with limited familiarity with mindfulness, more direct forms of feedback were considered to be appropriate. The following example from group 1 illustrates this notion:

\begin{quote}
\textit{ There are some apps they will record how long you focus or how long you spend time on meditation. But we don't want that because we think that will make users feel anxiety or nervous. So, we just want our device to teach our users how to breath} (G1\_M2)
\end{quote}

In addition, some participants imagined devices that provided more metaphorical or ambient forms of feedback for meditators with greater experience in mindfulness, aiming to subtly enhance their ongoing practice rather than directly guiding it. For example, G2\_H1 described a device that would warm a beverage on a cold day or cool it on a hot day, offering a gentle form of support that could be integrated seamlessly into every mindful moment.

Participants also emphasized the importance of designing devices that engage cognitive resources to an optimal extent, ensuring that no excess capacity remains available for unrelated thoughts or distractions.

\begin{quote}
\textit{Using this device, will occupy our whole brain, so we wouldn't have time to think about other stuff. I mean its difficult... Not difficult, but difficult enough to use all the capacity of the brain. So thinking about the temperature [device feedback] is not an easy thing, so you wouldn't think about your tomorrow exam when you are doing that [meditation with device] (G1\_M1)}
\end{quote}

\paragraph{Feedback that supports practice through subtle anchors, not metrics}

This theme illustrates participants' preferences on how the feedback should be presented. The overarching theme encompasses the sub-themes of reminders, anchors, tracking, and feedback qualities, each of which is examined in detail. In this context, reminders serve to prompt or encourage the initiation of meditation, whereas anchors function as supportive elements that facilitate and sustain the meditation.

Participants frequently highlighted \textit{reminders} as a common form of feedback. While some appreciated devices that gently prompted them to begin a session, many expressed a stronger preference for internally driven reminders rather than external cues. By this, participants referred to subtle forms of feedback that they could notice and interpret on their own, such as changes in breath, posture, or device-mediated signals, without requiring explicit notifications. Standard mobile app notifications such as ‘time to meditate' were often described as 'disruptive' and 'inauthentic', pushed by an algorithm. For example, G2\_H1 said, ``\textit{I really don't want to set a phone reminder because I have too much of that in real life anyway}''. This finding highlights the importance of leveraging users' agency, rather than issuing direct commands. As a solution, participants suggested that devices should be designed to foster autonomy and provide space for self-initiated engagement. They proposed \textit{non-nudging reminders} to build internally driven reminders. Here, non-nudging is a way to simply notice the presence of the device, which might serve as a subtle cue to initiate the meditation. 

However, a key challenge participants identified in designing such feedback was the \textit{intensity of reminders}. Signals that are too strong were described as jarring, while those that are too subtle risked being overlooked. Finding the right balance, tailored to individual preferences, emerged as an important design consideration. For example, Group 2 proposed a vibrating mat that prompts users to meditate. However, they also noted the over-saturation of vibrating devices in everyday life.
\begin{quote}
\textit{
Like, something that you can have your shoes off and maybe it's something quite comfy. Like, every so often it maybe, like, vibrates a bit. That one I'm less of a fan of just because, like, I think we have so many vibrating. Buzzing devices.}(G2\_H1)
\end{quote}

Participants also framed feedback as a form of an \emph{anchor}, providing cues that both inform the user and serve as a point of focus for meditation itself. Participants highlighted sensory stimuli, whether visual, auditory, or tactile, as anchors, and recognized the value in background cues that quietly support awareness without demanding it. They emphasized  considering sensory stimuli like tactile experiences and the human desire for touch could be a possible entry point for creating devices that make the user feel compassionate, grounded, and embodied. For example, Group 2 proposed a hug jacket, a wearable device that allows users to experience the sensation of self-hugging as a potential self-compassion device by emphasizing the touch-deprived nature of today's generation.

\begin{quote}
\textit{[Today] A lot of people are touch-poor or touch-deprived. You know, like they just don't have that real [physical] contact. So if you can get a garment that's like a hug. That could be really good for people who don't have that. Because a lot of people love touch. But they do not get the touch }(G2\_M1)
\end{quote}

\paragraph{Physical form of the device}

This theme expands on how participants envisioned the physical form and embodiment of mindfulness devices. Participants' discussed their preferences, design ideas and revealed how embodiment intersects with convenience, personal experience, and cultural associations with meditation practices.

Many participants highlighted the convenience of wearable devices. Wearables were seen as less intrusive, easier to integrate into daily routines, and capable of supporting mindfulness in real-time contexts. 

\begin{quote}
    \textit{So, the thing about clothing is like, if we're thinking about like daily reminders. Something that you wear, it's on you, all the time. I think the hard thing would be like if it's a device, like how do you transfer, like it would have to be transferrable to different garments. You know. Or it would have to be a garment that you wear, like a watch.} (G2\_H1)
\end{quote}

By being continuously accessible on the body, such devices could provide subtle feedback, through vibration, sound, or pressure without requiring users to actively engage with an external object. This form factor was often described as making mindfulness practices more seamless and less disruptive to other activities.

We identified that participants' personal experiences also influenced their ideas of how mindfulness devices should be embodied. For some, past experiences with fitness trackers or health monitoring devices shaped expectations around comfort, size, and functionality.

\begin{quote}
    \textit{
Temperature. So, like, an Apple Watch or something that takes your heart rate and you can set it to different heart rates warm up or cool down [...] like, a reminder (G3\_H1)}
\end{quote}

These reflections revealed how embodiment is not only a practical concern but also a deeply personal one, tied to individuals' previous encounters with technology and mindfulness.

\subsection{Learning from Focus Group Activity}


In this section, we discuss six findings that contrast with existing literature on designing tangible interactions for mindfulness contexts. First, participants consistently expressed a preference for wearable form factors, suggesting that wearables may be particularly well suited for integrating mindfulness practices seamlessly into everyday life. This finding aligns with our scoping review finding, \citet{liCodesigningMagicMachines2023}, who identified wearables and jewelry as a preferred form factors for enabling mindfulness practices that can be carried and used in diverse everyday contexts.

Second, our findings diverge from the trade-offs identified by \citet{liCodesigningMagicMachines2023} regarding tracking mindfulness outcomes versus supporting ongoing, process-oriented practice. While prior work explored balancing these approaches, our stakeholders outright rejected the notion of tracking mindfulness progress altogether. Participants frequently expressed concern that quantification and performance-oriented metrics conflict with the non-judgmental and experiential qualities central to mindfulness practice.

Third, participants' own definitions of mindfulness also extended beyond canonical formulations to include notions such as \textit{peacefulness}, \textit{joyfulness}, \textit{awareness}, and ``\textit{being present}'' which aligns with our scoping review finding, \citet{terzimehicReviewAmpAnalysis2019}'s work.

Fourth, several ideas drew explicitly from spiritual practices, particularly Buddhist traditions, reflecting how participants themselves framed mindfulness through culturally situated understandings rather than strictly clinical definitions. This contrasts with our literature review findings, which did not identify prior work engaging with spirituality practices and extend beyond to transcendence in mindfulness technologies. 

Fifth, our findings reinforce the broader opportunity to integrate mindfulness into everyday embodied activities should be socially and culturally resonant consistent with \cite{liMeditationUnderstandingEveryday2024a}. However, our participants expressed high interest in body-scanning practices, not only in seated meditation but also while lying down (\autoref{fig:design_space}d). Similarly, our scoping review identified examples such as mindful eating supported through sensory augmentation \cite{chenLivingBentoHeartbeatDriven2025}.

Sixth, our participants emphasized the importance of structure in mindfulness practice, resonating with prior findings of reminder and facilitator support in \cite{liCodesigningMagicMachines2023}. We observed that such structure operates at two levels: \emph{within-session} support and \emph{between-session} support. Within sessions, tangible devices can act as attentional facilitators \cite{liCodesigningMagicMachines2023}, supporting bodily awareness and sustained focus. One way this may occur is through what we term \textbf{sensory expansion}, where external augmentation of bodily signals enhances the salience of sensations that practitioners are encouraged to attend to, such as breathing. Participants described how vibrations synchronized with breathing could support concentration during focused attention meditation (FAM), echoing the externalized attentional support discussed in \citet{daudenroquetBodyMattersExploration2020}. Our scoping reviw found similar approaches that have been explored in systems such as ManifestingBreath ~\cite{farrallManifestingBreathEmpirical2023a} and Breathm~\cite{theodoreBreathmCalmDevice2024}, which aim to reduce attentional effort and promote relaxation.

Beyond attentional support, participants also highlighted opportunities for \textbf{sensory anchoring} (we term anchoring inspire by Mindful moments~\cite{tanMindfulMomentsExploring2023} working as ``acts as an anchor [P5]''), where tangible interactions intentionally evoke or sustain affective states during mindfulness practice. In particular, affective touch emerged as a promising direction. Participants described how textured materials and gentle pressure could evoke feelings of calmness, safety, and self-compassion. One illustrative proposal was a ``hug jacket'' (\autoref{fig:design_space}m), designed to simulate compassionate touch through pressure around the torso. Prior work demonstrates that affective touch can support anxiety reduction and physiological calming~\cite{zhaoAffectiveTouchImmediate2023}, intimate connection~\cite{cheokHuggyPajamaRemote2010, ozcanMultisensoryWearableBiofeedback2023}, and emotional communication~\cite{fooUserExpectationsMental2021} but we did not find work on mindfulness.

\section{Theoretical Framework For designing Tangible Devices for Mindfulness}
\label{sec:framework}
In this section, we consolidate our findings by extending a theoretical framework for HCI. Our novel theoretical framework serves both as an explanation (i.e. `explaining mechanisms') and an hypothesis building (i.e. `making predictions') instrument~\cite{imenda2014there}. In the following subsections we describe the framework's background, construction \& validation, and guide on how to use the framework, in accordance with the recommendations for HCI frameworks by~\citet{fang2026we}.

We propose two extensions of the \textbf{self-awareness, self-regulation, and self-transcendence (S-ART)} framework by~\citet{vagoSelfawarenessSelfregulationSelftranscendence2012}: 
\begin{itemize}
    \item Embodied Sensory Expansion (ESE) model (\autoref{fig:FA_framework}) extending concentrative practice for Focused Attention Meditation (FAM).
    \item Embodied Sensory Anchoring (ESA) model (\autoref{fig:OMM_framework}) extending open receptive practice for Open Monitoring Meditation (OMM).
\end{itemize}

\subsection{Background}

Researchers have proposed several frameworks to explain how mindfulness operates. Among the most influential is the Intention, Attention, and Attitude (IAA) model introduced by \citet{shapiroMechanismsMindfulness2006}, which builds upon Kabat-Zinn's definition of mindfulness as ``paying attention in a particular way: on purpose, in the present moment, and non-judgmentally'' \cite{kabat_zinnMindfulnessbasedInterventionsContext2003}. However, despite its conceptual significance, we found the IAA model difficult to operationalize within tangible device design due to its abstract and conceptually broad nature of constructs. In our work, we built upon (or ``adapt''~\cite{fang2026we}), \textbf{S-ART} framework by~\citet{vagoSelfawarenessSelfregulationSelftranscendence2012}. We chose this framework for three reasons. First, the framework conceptualizes mindfulness as a \textit{trainable skill} developed through three forms of meditation: focused attention meditation (FAM), open-monitoring meditation (OMM) and compassion/loving kindness meditation (CM/LKM). Second, the framework is grounded in established psychological and neurobiological findings. We believe that such grounding facilitates more explainable designs. For example, our framework enable hypothesis building on the interaction of the tangible system \& working memory and how it effects mindfulness. 
Finally, the framework inherently builds upon the mechanisms of mindfulness action proposed by \citet{holzelHowDoesMindfulness2011}. This theoretical continuity makes it \textit{more suitable} for validating our findings against the results of the scoping review, as both frameworks share compatible conceptual foundations regarding how mindfulness practices produce cognitive, neurobiological and affective changes.

In the following section, we provide a brief introduction to the \textit{concentrative practice} underlying FAM and the \textit{open monitoring receptive practice} underlying OMM, with relevance for HCI designers using simplified figures. For a comprehensive description of these models, their construction and theoretical foundations, see~\cite{vagoSelfawarenessSelfregulationSelftranscendence2012}.

\subsubsection{Concentrative Practice for FAM}

In ~\citet{vagoSelfawarenessSelfregulationSelftranscendence2012}'s model, the authors define ``\textit{focused attention practice involves sustained attention on a specific mental or sensory object: a repeated sound or mantra, an imagined or physical image, or specific viscerosomatic sensations. The object of focus can be anything, but the method described by the Satipattāna Sutta identifies a naturally occurring breath focus }''. 

The mechanism starts with the meditator's \textbf{Intention \& Motivation} initiating the practice by establishing a deliberate goal to sustain awareness on an \textbf{Intended Object (e.g. Breath)}. Which is guided by the \textbf{Set Formation (Practice Instructions)} held in \textbf{Working Memory}. Then the practitioner recruits \textbf{Focused Attention} and attentional \textbf{Orienting} mechanisms to achieve \textbf{Mental Stabilization}, continuously engaging with the breath while maintaining meditation instructions in awareness (see Attention Loop in \autoref{fig:FA_framework}).

Inevitably, an \textbf{Unintended Object (e.g. sensory or mental event)} could emerge as a distraction, eliciting an \textbf{Affective Response (which could be positive, negative or neutral)} and recursive \textbf{Mental Proliferation} through activation of \textbf{Episodic \& Procedural Memory} events. Then the practitioner detects their mind wandering through \textbf{Executive Monitoring (meta-awareness)} and initiates \textbf{Decentering}, recognizing thoughts and emotions as transient mental events rather than identifying with them as self. This facilitates \textbf{Response Inhibition} and \textbf{Emotion Regulation}, cultivating \textbf{Equanimity} and reducing affective reactivity, rumination, sympathetic arousal, and cognitive elaboration. The practitioner then disengages from distraction and reorients attention toward the breath (see Distraction Loop in \autoref{fig:FA_framework}). When repeating this process over time the attentional control becomes increasingly automatic through \textbf{Motor Learning}. Also, the required \textbf{Effort} progressively decreases, resulting in stabilized attentional regulation, and enhanced meta-awareness  (see Long-term Loop in \autoref{fig:FA_framework}).

\subsubsection{Process Model of Open Monitoring Receptive Practice for OMM}
In contrast to concentrative practice,~\citet{vagoSelfawarenessSelfregulationSelftranscendence2012}'s open monitoring receptive practice model replaces Focused Attention on a single intended object (e.g. breath) with diffused \textbf{Ambient Attention} directed toward the continuous flow of experience without a fixed object of focus. The meditator flexibly monitors arising and passing \textit{phenomena} across \textbf{Auditory/Linguistic}, \textbf{Visual}, and \textbf{Viscero-Somatic} domains through \textbf{Mental Noting \& Labeling}. This mental noting and labeling process leads to an \textbf{Affective Response} (positive, negative, or neural) by invoking \textbf{Episodic \& Procedural Memory} associated with them (see Mental Noting/Affective Loops in \autoref{fig:OMM_framework}). 

Similar to FAM, the practitioner detects mind wandering through \textbf{Executive Monitoring (meta-awareness)} and initiates \textbf{Decentering}, recognizing thoughts and emotions as transient mental events rather than identifying with them as self. This facilitates \textbf{Response Inhibition} and \textbf{Emotion Regulation}, cultivating \textbf{Equanimity} and reducing affective reactivity, rumination, sympathetic arousal, and cognitive elaboration. The practitioner then disengages from distraction and reorients attention toward the breath. When repeating this process over time the attentional control becomes increasingly automatic through \textbf{Motor Learning}. Also the required \textbf{Effort} progressively decreases, resulting in stabilized emotional regulation, enhanced meta-awareness, and reduced habitual self-referential processing (see Long-term learning Loop in \autoref{fig:OMM_framework}).


\subsection{Framework: Construction \& Validation}

Our framework was developed iteratively and validated through a two-stage process. First, the literature identified through the scoping review was randomly divided into two subsets. Using the first subset, three authors collaboratively developed the framework by progressively incorporating additional components into the S-ART framework~\cite{vagoSelfawarenessSelfregulationSelftranscendence2012} to account for the mechanisms of action identified in the scoping review. Subsequently, the first author presented the preliminary framework to two additional authors who had not been involved in its development to evaluate the \textit{interpretability of designs using the framework}. These authors were asked to interpret and explain the mechanisms reported in the second subset of papers using the proposed framework. Based on the feedback, we updated the framework to explicitly represent controllable and measurable (through psychological, physiological or behavioral markers) components from a design-oriented perspective.

Finally, the framework was validated with three independent HCI designers, each possessing more than three years of professional experience and no prior involvement in the project. The designers were asked to use the framework to rapidly generate hypotheses. We report some examples of the hypotheses generated during this process in the \hyperref[sec:how_to_use_the_model]{How to use} section.


\subsection{Embodied Sensory Expansion Model for FAM}

\begin{figure*}[t]
    \centering
    \includegraphics[width=1\linewidth]{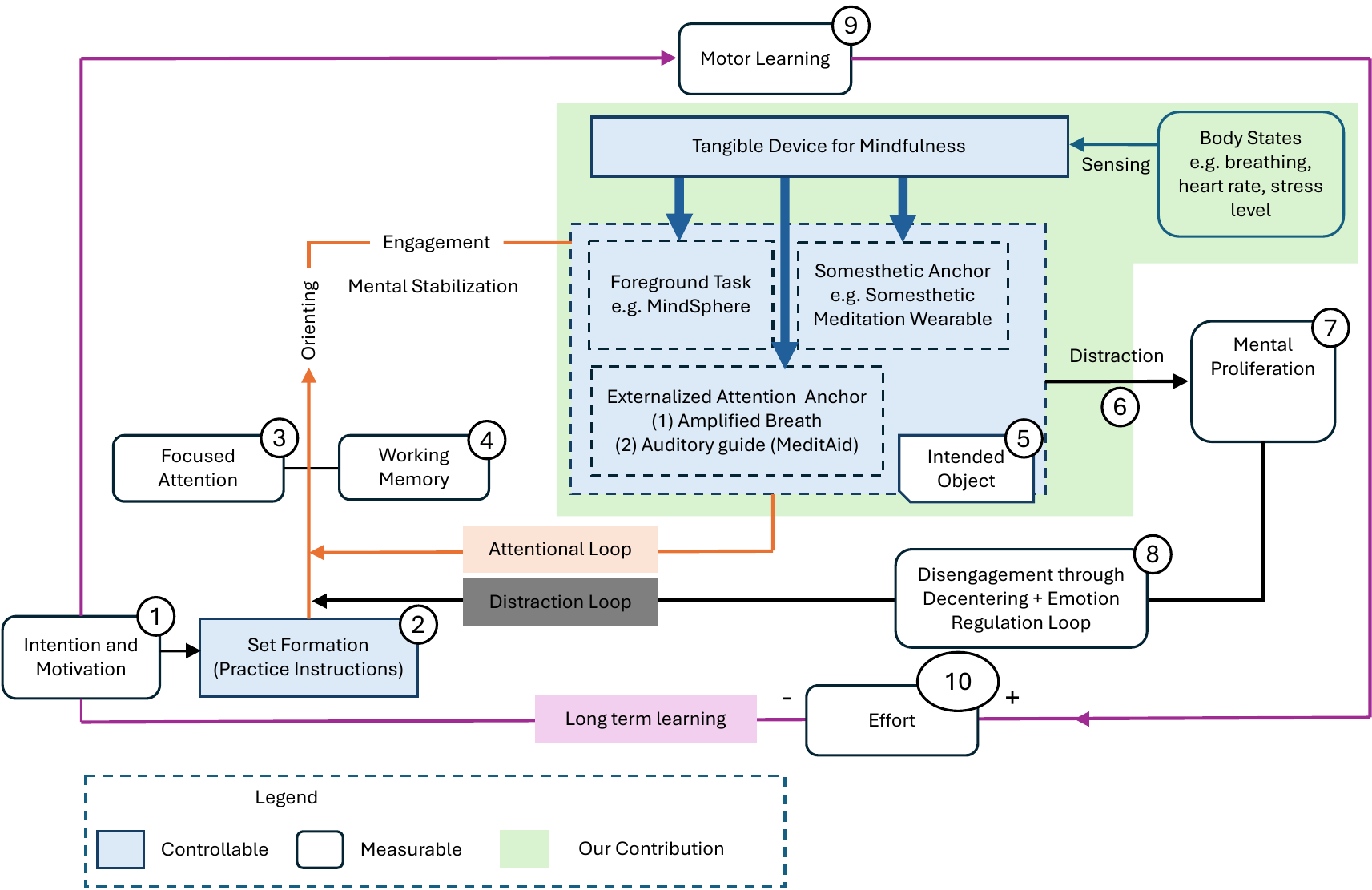}
    \caption{The \textit{embodied sensory expansion model}, adapted from the simplified concentrative practice model for Focused Attention Meditation proposed by~\cite{vagoSelfawarenessSelfregulationSelftranscendence2012}. The framework comprises three interconnected loops: (1) the basic attentional loop; (2) the distraction loop, in which distraction leads to mental proliferation, emotion regulation, and subsequent return to mental stabilization; and (3) the long-term learning loop, through which repeated practice strengthens attentional regulation via motor learning, thereby reducing the effort required to sustain focused attention.}
    \Description{The figure shows the proposed conceptual framework for sensory expansion framework}
    \label{fig:FA_framework}
\end{figure*}

Conceptually, the model preserves the overall attentional loop described in concentrative process model of S-ART Framework~\cite{vagoSelfawarenessSelfregulationSelftranscendence2012}: intention formation, attentional stabilization, distraction, meta-awareness, disengagement, and re-orientation. However, it introduces a new mediating layer: the \textit{Tangible Device for Mindfulness}, extends the role of the intended object (see \autoref{fig:FA_framework}). In doing so, tangible devices can be used to externalize, scaffold, amplify, or dynamically regulate attention through embodied interaction and sensing-supported feedback. The detailed mechanism is explained below.

Similar to the original concentrative model, the practitioner establishes a deliberate intention to engage in mindful activity by forming \textbf{Intention and Motivation} (\autoref{fig:FA_framework} \circref{1}). Following intention formation, the practitioner establishes a \textbf{Set Formation} (\autoref{fig:FA_framework} \circref{2}) phase, where practice instructions and \textit{interaction rules} are defined. In conventional meditation this may consist of instructions such as ``attend to the breath'' or ``return attention when distracted.'' However, within our framework, the set additionally includes the interaction logic of the tangible system, becoming part of the attentional set architecture. For example, the system may instruct users to synchronize with rhythmic haptic breathing cues~\cite{macik2017breathinga}, attend to thermal stimulation patterns~\cite{dublinJourneyInwardSomaesthetic2024} and sustain smooth motor interaction~\cite{feijs2005design}. Once the executive set is established, \textbf{Focused Attention} (\autoref{fig:FA_framework} \circref{3}) and \textbf{Working Memory} (\autoref{fig:FA_framework} \circref{4}) processes support \textit{Mental Stabilization}. However, unlike traditional mindfulness practice, our framework proposes that the \textbf{Intended Object}  (\autoref{fig:FA_framework} \circref{5}) can be supported through the tangible device. This could take three forms and leads to the basic \textit{Attention Loop} in \autoref{fig:FA_framework}.

One major mechanism involves the \textit{Externalized Attentional Anchors}, where the device provides persistent temporally structured cues that support attentional engagement. As discussed in \hyperref[para:attention_reg_externalized_attention_anchors]{Paragraph 3.2.1.1}, multiple systems operationalize this through breathing entrainment. For example, Mindful Moments~\cite{tanMindfulMomentsExploring2023} externalizes respiratory pacing through visual cues, while the handheld breathing device by~\citet{macik2017breathinga} transforms breathing into rhythmic haptic inflation and deflation. Across these systems, the attentional object is a dynamically perceived sensory output produced by the device which functions as an external attentional stabilizer that reduces attentional drift and lowers the cognitive effort required to maintain focus.

Second, a related but distinct mechanism involves \textit{Somaesthetic Attention Anchors} (see \hyperref[para:attention_reg_somaesthetic_attention_anchors]{Paragraph 3.2.1.3}), where attention is grounded through enhanced bodily sensation rather than temporally structured pacing. Systems such as the somaesthetic meditation wearable~\cite{ezerSomaestheticMeditationWearable2024a} and the walking meditation mat~\cite{dublinWalkingMeditationMat2025a} guide attention through thermal stimulation distributed across the body.

Within the framework, \textbf{Distraction} (\autoref{fig:FA_framework} \circref{6}) remains inevitable. External sensory events, intrusive thoughts, emotional reactions, or unrelated mental activity continue to compete for attentional resources
This leads to \textit{Distraction Loop}, which includes \textbf{Mental Proliferation} (\autoref{fig:FA_framework} \circref{7}), and then disengagement through meta-awareness and emotion regulation (\autoref{fig:FA_framework} \circref{8}). Yet unlike traditional meditation where re-orientation depends entirely on internally generated meta-awareness, tangible systems may actively support the process by incorporating neuro-- or bio--sensing.

This occurs particularly through neuro-- or bio--sensed \textit{Closed-Loop Systems} (see \hyperref[para:attention_reg_closed_loop_systems]{Paragraph 3.2.1.2}), where sensing technologies dynamically infer attentional or physiological states and adapt feedback accordingly. Systems such as meditAid~\cite{sas2015meditaid} use EEG signals to detect fluctuations in attentional focus and modify auditory entrainment in response. Similarly, \cite{paredesJustBreatheIncar2018} detect stress levels and provide adaptive haptic breathing interventions during driving. In these cases, the tangible device supports directly in attentional regulation by detecting signs of dysregulation and dynamically adapting the feedback. We believe such regulation, stem from improving the signal-to-noise ratio of the \textbf{intended object}. Behavioral closed-loop systems similarly operationalize distraction through observable deviations from desired behavior. For example, \cite{pryssPersonalizedSensorSupport2018} monitor walking speed and intervene when users deviate from a mindful pace, while SWAN~\cite{khotSWANDesigningCompanion2020} interrupts distracted eating behaviors through a mechanized spoon. These systems conceptualize attentional disruption behaviorally and support re-engagement through corrective interaction.

Finally, framework incorporates \textit{Task-Coupled Attentional Constraints} (see \hyperref[para:attention_reg_task_coupled]{Paragraph 3.2.1.4}), where mindfulness emerges directly through task engagement and sensorimotor coordination. In MindSpheres~\cite{feijs2005design}, users maintain smooth coordinated motion of handheld spheres, while feedback reflects the quality of ongoing interaction. Similarly, embodied appliance interactions proposed by \cite{Engagement_through_Embodiment_16} require continuous bodily modulation to sustain operation. In these systems, the \textbf{Intended Object}  (\autoref{fig:FA_framework} \circref{3}) is not externally represented breathing or biofeedback, but the unfolding embodied interaction itself. Attention regulation therefore emerges from sustained sensorimotor coupling.

Similar to concentrative practice by \citet{vagoSelfawarenessSelfregulationSelftranscendence2012}, when repeating this process over time the attentional control becomes increasingly automatic through \textbf{Motor Learning} (\autoref{fig:FA_framework} \circref{9}). Also the required \textbf{Effort} (\autoref{fig:FA_framework} \circref{10}) progressively decreases, resulting in stabilized emotional regulation, enhanced meta-awareness, and reduced habitual self-referential processing (see \textit{Long-term learning loop}).

\begin{figure*}[t]
    \centering
    \includegraphics[width=1\linewidth]{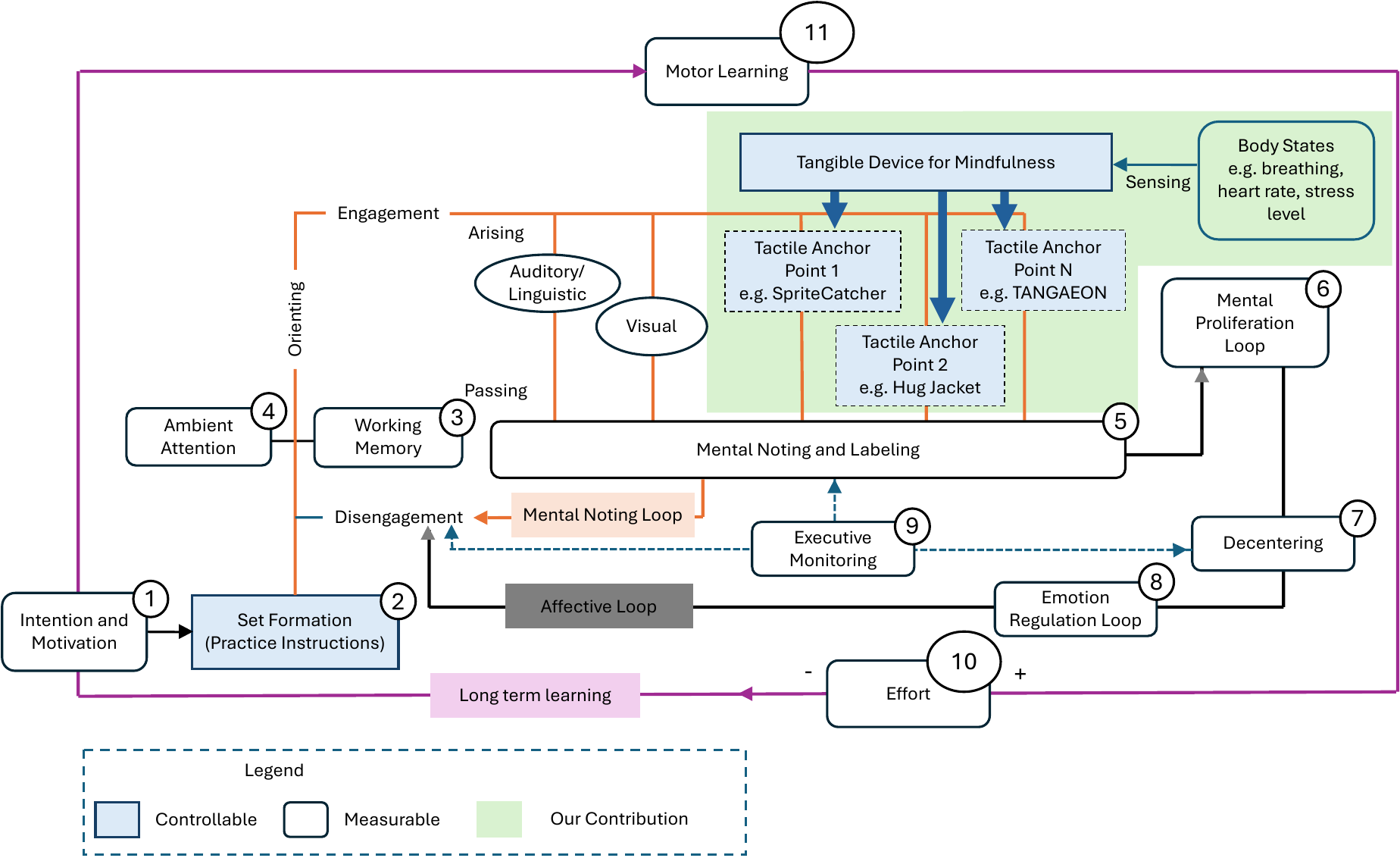}
    \caption{The \textit{embodied sensory anchoring model}, adapted from the simplified open receptive practice model for Open Monitoring Meditation proposed by~\cite{vagoSelfawarenessSelfregulationSelftranscendence2012}. The framework comprises three interconnected loops: (1) the basic mental noting loop; (2) a distraction loop in which mental noting and labeling evoke affective responses that lead to mental proliferation, emotion regulation, and a subsequent return new arising phenomena; and (3) a long-term learning loop through which repeated practice strengthens open receptiveness via motor learning, reducing the effort required to sustain receptive awareness.}
    \Description{The figure shows the proposed conceptual frameworks for embodied tangible design}
    \label{fig:OMM_framework}
\end{figure*}

\subsection{Embodied Sensory Anchoring Model for OMM}

Conceptually, the model preserves the overall receptive cycle described in open monitoring receptive process model of S-ART framework~\cite{vagoSelfawarenessSelfregulationSelftranscendence2012}: intention \& motivation, set formation, ambient attention, monitoring of arising experience, mental noting and labeling, disengagement through decentering and emotion regulation, and eventual non-reactive awareness. However, it introduces a new mediating layer: the \textit{Tangible Device for Mindfulness}, helping during mental noting \& labeling (see \autoref{fig:OMM_framework}). In doing so, we propose that tangible devices can support at self-referential proliferation, decentering and positive affective escalation.

Consistent with the original OM model, practice begins through \textbf{Intention and Motivation} (\autoref{fig:OMM_framework} \circref{1}), where users deliberately engage in mindful observation. This is followed by \textbf{Set Formation} (\autoref{fig:OMM_framework} \circref{2}) in \textbf{Working Memory} (\autoref{fig:OMM_framework} \circref{3}),  often involving instructions such as ``\textit{observe whatever arises},'' ``\textit{notice thoughts without attachment},'' or ``\textit{maintain open awareness}.'' Within our framework, the set formation (i.e. practice instructions) additionally incorporates the sensing and interaction structure of the tangible system itself. For example, users may be instructed to observe/interact-with the soft-textile that can elicit positive affective response (similar to Coral Morph~\cite{huangCoralMorphArtistic2025}). However, unlike concentrative practice, where attention narrows toward a single intended object, OM recruits \textbf{Ambient Attention} (\autoref{fig:OMM_framework} \circref{4}), supporting broad receptive awareness toward arising and passing phenomena~\cite{vagoSelfawarenessSelfregulationSelftranscendence2012}. Our framework proposes that such arising and passing phenomena can be supported to employ mechanisms of action of mindfulness through the tangible device using \textit{tactile anchor points} to elicit emotions which helps with \textbf{Mental Noting and Labeling} (\autoref{fig:OMM_framework} \circref{5}). This could take several forms.


First, when the sensing technologies are coupled with open-monitoring meditation, they may support body awareness. This could be employed with few strategies. The first strategy, \textit{Embodied Amplification} (see \hyperref[para:body_awarness_embodied_amp]{Paragraph 3.2.2.1}), increases perceptual access to subtle physiological processes by transforming them into directly perceivable sensory experiences. Systems such as PiHearts~\cite{aslanPiHeartsResonatingExperiences2020}, ambienBeat~\cite{choiAmbienBeatWristwornMobile2020a}, and Manifesting Breath~\cite{farrallManifestingBreathEmpirical2023a} physically render heartbeat or breathing through haptic, pneumatic, or tangible movement. Similar to externalized attentional anchors, these systems externalize otherwise inaccessible internal states. However, unlike attentional entrainment systems whose primary goal is stabilization, embodied amplification systems foreground interoceptive awareness itself.

The second strategy, \textit{Externalized Biofeedback Representations} (see \hyperref[para:body_awarness_externalized_biofeedback]{Paragraph 3.2.2.2}), transforms physiological processes into environmental perceptual phenomena. Systems such as \cite{vidyarthi2012sonic}, MoodWings~\cite{MoodWings2013}, Wisp~\cite{gamboaWispDronesCompanions2023}, and Coral Morph~\cite{huangCoralMorphArtistic2025} relocate internal physiological states into visual, auditory, kinetic, or ambient representations. In these cases, body awareness emerges through observing one's physiology as an external object.

Finally, \textit{ Scaffolded Somatic Attention } (see \hyperref[para:body_awarness_scaffolded_somatic_attention]{Paragraph 3.2.2.3}) structures how attention traverses bodily processes. Systems such as Soma Mat~\cite{stahlSomaMatBreathing2016} and \cite{kuDisImmersionMindfulness2023} guide awareness spatially across bodily regions using thermal or auditory cues, while movement-based systems such as \cite{tenbhomerDesigningPersonalizedMovementbased2018} use EMG sensing to support awareness of muscular activation. In these systems, the device does not merely amplify bodily signals but organizes the attentional trajectory itself.

The framework further proposes that tangible systems can support \textbf{Mental Proliferation} (\autoref{fig:OMM_framework} \circref{6}) \& \textbf{Emotion Regulation} (\autoref{fig:OMM_framework} \circref{8}) by modifying the user's relationship to emotional experience (see \hyperref[sec:emotion_regulation]{\textit{Emotion Regulation}}). This may occur either through reinterpretation of emotional states or through sustained non-reactive exposure to affective experience. For example, The Drop the Beat system~\cite{seolDropBeatVirtual2017} allowing users to observe and physically manipulate a virtual representation of their heartbeat during panic-inducing scenarios. Similarly, the Breathing Scarf~\cite{cochraneBreathingScarfUsing2022} combines biofeedback-guided breathing regulation with reflective journaling supporting reinterpreting emotional experience through embodied sense-making. Also Sprite Catcher~\cite{Sprite_Catcher_2017} uses a disruption and redirection systems where device intentionally interrupts rumination. This closely parallels the disengagement and response inhibition processes described in the original concentrative process model, where meta-awareness interrupts proliferative cognition and redirects attention toward the intended object.

Within the OM framework, these systems support response inhibition and emotional equanimity by reducing reactive escalation and promoting sustained observational awareness toward affective states.

Also the framework proposes that tangible systems facilitate \textbf{Decentering} (\autoref{fig:OMM_framework} \circref{7}), hence change in perspective on the self (see \hyperref[sec:change_in_perspective]{\textit{Change in Perspective on the Self}}), where thoughts, emotions, and bodily states are experienced as transient events rather than reflections of the self. A prominent strategy involves metaphorical and environmental externalization. For example, unlike direct biofeedback systems, Inner Garden system~\cite{rooInnerGardenConnecting2017a} introduces abstraction that supports self-reflective observation.


\subsection{Framework: Methodological Constraint For Compassion-based Meditation}

In the S-ART framework proposed by~\citet{vagoSelfawarenessSelfregulationSelftranscendence2012}, a third category of contemplative practice, Ethical Enhancement (EE), is introduced to account for practices such as loving-kindness meditation (LKM) and compassion meditation (CM). Importantly, compassion-based practices have been characterized as EE-style practices only when accompanied by the concurrent cultivation of open presence; without such cultivation, the practice has been argued to be less effective~\cite{lutz2007meditation, vagoSelfawarenessSelfregulationSelftranscendence2012}. Accordingly, the EE framework described by \citet{vagoSelfawarenessSelfregulationSelftranscendence2012} already incorporates an Open Monitoring (OM)  component, which serves to facilitate awareness of any arising modality of experience (see the OM block embedded within the EE framework in \cite{vagoSelfawarenessSelfregulationSelftranscendence2012}). Therefore, in the present work, we do not propose a separate framework for EE practice. Instead, we consider the OM component embedded within EE practice to be replaced by our updated OM framework.

\section{Discussion and Research Agenda}
\label{sec:discussion}

Our purpose in writing this paper was to make sense of the rapidly growing, inherently interdisciplinary field of tangible devices for mindfulness. One key observation emerging from the reviewed literature is that the majority of existing systems primarily target attentional regulation and body awareness, leaving other dimensions of mindfulness practice, such as compassion, non-judgment, and emotional regulation, comparatively underexplored. We argue that this narrowing of focus may partly stem from the absence of an integrative theoretical framework that explains how tangible systems support distinct mindfulness processes and outcomes.

In response, we develop a theory-driven framework comprising two complementary models: \textbf{Embodied Sensory Expansion}, targeting concentration practice in Focused Attention Meditation (FAM), and \textbf{Embodied Sensory Anchoring}, targeting open-receptive practice in Open-Monitoring Meditation (OMM). We expect this framework to support the HCI community in three ways: first, by enabling stronger hypothesis building and explanatory power for understanding how and why tangible mindfulness systems work through a neurobiological grounding; second, by revealing research gaps through positioning prior work within a coherent theoretical structure and guiding future research directions; and third, by establishing a principled basis for evaluation through clarifying what should be measured to determine whether a tangible device achieves its intended mindfulness outcomes.


In the following subsections, we aim to demonstrate how the framework can be used to shape future research and practice. In the \textit{How to Use} section, we illustrate how the framework can support hypothesis generation, explain system behavior, sharpen theoretical reasoning about tangible mindfulness interventions, and identify research gaps. In the \textit{How to Evaluate} section, we present recommendations, offering concrete guidance for assessing whether tangible mindfulness devices achieve their intended goals, while also identifying under-explored areas that warrant further empirical research.

\subsection{How to use: Implications From the Framework}
\label{sec:how_to_use_the_model}

Across both models, the primary controllable elements are the \textbf{tangible device} and the \textbf{practice instructions} (or set formation). All other components of the framework can be treated as dependent or mediating variables, including attentional stability, ambient attention, working memory, body awareness, meta-awareness, emotion regulation, decentering, motor learning, perceived effort, and habit formation (see elements in \autoref{fig:FA_framework} and \autoref{fig:OMM_framework}). In the following sections, we discuss how each controllable element may be operationalized across both FAM, OMM, and CM practices and highlight the research gaps.


\subsubsection{Research Gaps--Tangible Device Perspective}

First, the framework proposes that \textbf{tangible devices} can be systematically varied to examine how different forms of technological mediation influence mindfulness processes. Relevant device-level variables include \textbf{embodiment, sensing, feedback, and adaptivity}. 

\paragraph{Research Gaps--Embodied Sensory Expansion model} 
Based on the \textbf{Embodied Sensory Expansion} model, tangible devices may support \textit{focused attention meditation} across three interrelated loops. First, within the \textbf{Attentional Loop} (see \autoref{fig:FA_framework}), tangible devices strengthen focused attention by increasing the signal-to-noise ratio of the intended object of meditation. Prior quantitative studies have demonstrated that visual~\cite{tanMindfulMomentsExploring2023} and haptic~\cite{paredes2017evaluating} feedback can support concentration and attentional stability during meditation practice. In contrast, existing work on other modalities has primarily reported qualitative findings. For example, \citet{dublinWalkingMeditationMat2025a} suggested that thermal feedback may support attentional engagement, while \citet{wang2024design} explored a full-body wearable embodiment approach. However, further quantitative empirical evaluation would help strengthen the understanding of the effectiveness of these approaches. 

Furthermore, our framework proposes a relationship between working memory and tangible systems in mindfulness practice. Specifically, we posit that the tangible embodiment of the intended object of meditation may occupy greater working memory resources, reducing the cognitive resources available for unintended objects, which may contribute to distraction. This proposition emerged qualitatively from both our consultation activity and findings from the scoping review (see \cite{fooSoftRoboticCompression2020}). However, this relationship remains largely theoretical, and further quantitative empirical evaluation is needed to strengthen the understanding of how tangible interactions intersect with working memory processes in mindfulness practice.

Considering the \textbf{Distraction Loop} (see \autoref{fig:FA_framework}), tangible devices support practitioners during distraction and mind wandering by intervening in processes associated with mental proliferation, affective reactivity, and decentering by assisting users in recognizing distraction. For example,~\citet{sas2015meditaid} used EEG sensing to detect fluctuations in attention and adapted auditory feedback when distraction was detected. While this demonstrates the feasibility of adaptive neuro-feedback systems, other feedback modalities and forms of embodiment remain underexplored. Also, investigating less intrusive approaches for detecting mind wandering (i.e., distraction) may help broaden the applicability and accessibility, reducing the burden associated with wearable sensing technologies.

Considering the \textbf{Long-term loop }(see \autoref{fig:FA_framework}), the relationship between tangible devices, effort reduction, and motor learning remains largely unexplored. We claim the meditative effort to decrease with practice duration, reflecting a progressive increase in the motor-learning. To date, this longitudinal relationship between tangible device support and motor learning has received little empirical attention and remains a significant research gap.

Also, as proposed in our model for FAM, the primary purpose of the tangible device is to support ongoing attention regulation (i.e., attention-distraction loops) and the training of attentional control over time (i.e., the long-term loop) by fostering `meta-awareness'. Therefore, a successful intervention using the device would be expected to improve attention-related measures compared to conventional meditation practices. We also anticipate observing changes in affective responses. Based on our framework, we claim that during the intervention period, users spend less time engaged in the distraction loop (i.e., which includes the affective response) and more time within the attention regulation loop. From the user's perspective, this may manifest as improvements in emotion regulation and may yield mixed findings in affective response related measures~\cite{miriEvaluatingPersonalizableInconspicuous2020, choMindfulTouchMidair2025, tanMindfulMomentsExploring2023}. However, we do not expect such interventions (i.e., attention training) to develop `mindful' emotion regulation as a skill through reappraisal, extinction, and exposure.


\paragraph{Research Gaps--Embodied Sensory Anchoring model} 
Based on the \textbf{Embodied Sensory Anchoring} model, tangible devices support \textit{open monitoring meditation} across three interrelated loops. Within the \textbf{Affective Loop} (see \autoref{fig:OMM_framework}) tangible devices can support--disengagement from mental proliferation, affective response, decentering, and emotion regulation. First, prior work in affect regulation has investigated how tangible interactions can elicit positive affect and support the logging and externalization of affective responses~\cite{guribye2016designing, zhouTangibleAffectLiterature2024}. This raises the question: \textbf{can tangible interaction be intentionally designed to support emotional reappraisal during mindfulness practice?} Several tangible systems have been proposed in this direction~\cite{huttonReMiNDImprovingEmotional2019, cochraneBreathingScarfUsing2022}; however, their efficacy of these systems remains insufficiently explored through empirical evaluation. 


Also, the framework proposes that tangible devices may intentionally evoke negative affective emotions as part of mindfulness and emotional regulation practices. However, for such experiences to remain constructive, we argue that they should occur under controlled conditions. For instance, in SpriteCatcher~\cite{Sprite_Catcher_2017}, the device is designed to capture negative emotions (i.e., ``catch sprites'') and subsequently replay them (i.e., ``release the sprites'') in the presence of a therapist and under professional supervision. This approach enables users to safely revisit and process difficult emotional experiences, transforming negative affect into opportunities for reflection and therapeutic engagement in a `mindful' manner (i.e., non-judgmental). However, the efficacy of such a system remains insufficiently explored through empirical evaluation.

Considering the \textbf{Long-term Learning Loop}, the relationship between tangible devices, effort reduction, and motor learning also remains underexplored. As discussed by \citet{vagoSelfawarenessSelfregulationSelftranscendence2012}, meditative effort is hypothesized to decrease with practice duration, reflecting a progressive reduction in habitual mental proliferation and rumination. To date, this longitudinal relationship between tangible device support and motor learning in mindfulness has received little empirical attention and remains a significant research gap.


\subsubsection{Research Gaps--Tangible Device-Practice Instruction Interaction Perspective}

From the \citet{vagoSelfawarenessSelfregulationSelftranscendence2012}'s framework we identifies \textbf{practice instructions} as a distinct source of variation that shapes how the device is cognitively and behaviorally engaged during practice~\cite{ SupportingCognitiveReappraisal2024, NETO2025103459, TTSandEmbodiments2023}. For example, ~\citet{menhart2022effects} examined how the type of voice--human male/female and synthetic male/female--can impact users' levels of relaxation, perceived usefulness, and enjoyment when following a guided meditation, ~\citet{silvestre2023metaphor} examined how the phrasing of meditation language affects engagement, and ~\citet{nuttall2025linguistic} examined which linguistic metaphors are commonly employed within guided meditation practices. 

Building upon this, our theoretical framework enables the formulation of new hypotheses regarding the potential impact of tangible devices on mindfulness, particularly concerning the \textbf{interaction effects between practice instructions and tangible devices}. We propose that the ways in which practitioners are instructed to \textbf{interact with a device and interpret its feedback} may substantially influence the efficacy of the interaction. For example, our scoping review identified evidence suggesting such interaction effects. \citet{EncouragingBreath2026} reported that, over a six-week intervention, participants in a device+audio guidance condition demonstrated greater improvements in body awareness, mind--body connection, and acceptance of feelings and bodily sensations compared to an audio-only condition. However, the underlying mechanisms through which instructional framing and tangible interactions jointly shape mindfulness outcomes remain underexplored. 

For example, an underexplored question concerns how identical devices may produce different mindfulness outcomes under varying instructional framing, both \textit{between sessions} and \textit{within a session}. \textit{Between sessions}, the extent to which the same tangible system supports different mindfulness processes under FA versus OM instructions remains insufficiently understood. For example, a tangible device with neuro-feedback framed for mind-wandering may encourage `meta-awareness' during FA, whereas the same device framed as a negative emotion detector (e.g., see catching sprites in \cite{Sprite_Catcher_2017})) aid may instead support non-judgmental observation of negative thoughts. \textit{Within a session}, it remains unclear how linguistic cues (instructions) and embodied metaphors (for example, see \cite{daudenroquetInteroceptiveInteractionEmbodied2021a}) alter users' interpretation of tangible interactions and biofeedback representations.

\subsection{How to evaluate: Evaluating Devices For Mindfulness}

To start, a notable observation is that the majority of mindfulness devices have been developed within the broader field of affect computing, and only a few studies have reported validated psychological measures of mindfulness. Many rely on physiological proxy measures, such as HRV-related metrics, to infer mindfulness outcomes. While these measures may capture relaxation, they do not necessarily reflect the broader psychological dimensions associated with mindfulness. For example, mindfulness practice does not always lead to relaxation~\cite{lubertoPerspectiveSimilaritiesDifferences2020}; constructs such as \textit{decentering}--the ability to observe one's own thoughts with detachment--cannot be adequately captured through physiological measures. Therefore, we recommend that future research should more explicitly integrate psychological constructs of mindfulness into evaluation frameworks. For broader reviews of validated mindfulness scales, readers may refer to~\cite{sauer2013assessment, bergomi2013assessment, chems2026measuring}. However, it is important to note that these reviews were primarily conducted around 2013 \cite{bergomi2013assessment, sauer2013assessment} and may not reflect more recent developments in mindfulness measurement, including newer instruments such as the \textit{Embodied Mindfulness Questionnaire~\cite{khoury2023embodied}}. Furthermore, some scales may be particularly relevant to the HCI community despite receiving limited attention in the above reviews. For example, the \textit{State Mindfulness Scale (SMS)} was excluded in~\cite{chems2026measuring} due to the limited body of work available at the time. However, the SMS may be particularly relevant for the HCI community as it includes items assessing both \textit{mind}-related awareness (e.g., I was aware of the thoughts that came up in my mind'') and \textit{body}-related awareness (e.g., I noticed physical sensations come and go''), making it potentially valuable for studies of embodied interaction.

Second, when evaluating mindfulness interventions, we recommend that the outcome measures should align with the intervention duration and goals. For brief single-session studies involving tangible mindfulness devices, state mindfulness measures may be more appropriate than trait measures, as they are more sensitive to immediate changes following the intervention. Nevertheless, reporting baseline trait mindfulness as a covariate or participant characteristic remains important, as individual mindfulness dispositions may shape responses to mindfulness technologies. Prior research has shown that trait mindfulness is positively associated with pleasant affect~\cite{brown2003benefits} and agreeableness~\cite{thompsonEverydayMindfulnessMindfulness2007}, and negatively associated with depression~\cite{cashWhatFacetsMindfulness2010} and social anxiety~\cite{brown2003benefits}. Accounting for these individual differences may improve the interpretation and comparability of findings across studies.

Third, beyond mindfulness, we recommend tangible devices to support broader emotion-regulation processes (i.e., the distraction loop in ESE and the affective loop in ESA). Consequently, evaluations may benefit from incorporating measures related to anxiety, stress, or emotional well-being as secondary outcomes. However, researchers should also remain mindful of participant burden when selecting such measures. For example, although the State-Trait Anxiety Inventory (STAI) is widely used, using the shorter validated versions may reduce fatigue and improve participant engagement without compromising measurement quality~\cite{tluczek2009support}. This may be particularly important in studies involving repeated measures or lengthy survey batteries.

Fourth, we found that a substantial proportion of the reviewed papers relied on custom questionnaires to capture embodied, experiential, and interaction-specific aspects of engaging with tangible mindfulness devices (see Section~3.3 -- Psychological, Cognitive and Behavioral Evaluation). This highlights the absence of standardized, validated measures for evaluating embodied mindfulness interactions. Future research would benefit from the development of validated questionnaires that are sufficiently sensitive to detect subtle changes during short-term interventions (e.g. 10--15 minute interventions commonly used in existing studies). The development of such a validated instrument would improve reliability and construct validity, enabling stronger comparability, replication, and synthesis of findings across studies.

Fifth, we identified four primary evaluation dimensions across the reviewed studies: psychological, physiological and behavioral outcomes, usability, and subjective experience. However, only a limited number of studies incorporated measures spanning all four dimensions. As a result, many evaluations provided only a partial understanding of the effectiveness of tangible mindfulness devices. To achieve a more comprehensive assessment, we recommend that future research adopt multidimensional evaluation approaches that integrate all four aspects.

Finally, our framework proposes an additional long-term intervention component through the long-term learning loops in both frameworks. Based on the proposed framework, we suggest that long-term effects may be reflected in reduced effort required to sustain practice, increased motor learning and automaticity, and stronger intention and motivation to meditate (e.g., observed through increased meditation frequency over time). However, we also acknowledge that participants' mindfulness practice needs to be sustained without technological support (from the consultation activity). Therefore, we recommend that future research explicitly consider and model users' dependency over time, ensuring that reliance decreases while mindfulness skills increase.

\subsection{Limitations}
Our work has several limitations that should be acknowledged. First, in \citet{terzimehicReviewAmpAnalysis2019}' review, the design space for mindfulness technologies is categorized into three broad areas: (1) mobile applications, (2) extended reality (XR), and (3) tangible systems. In developing our framework, we focused exclusively on tangible devices and did not consider the broader literature on mobile applications and XR-based mindfulness interventions. We made this decision as tangible systems often involve purpose-built hardware and software, providing designers with greater flexibility for embodied and context-specific interactions. However, this focus limits the generalizability of our framework. Future work should therefore investigate whether and how our framework can be adapted to mobile and XR-based mindfulness technologies.

Second, during our scoping review, we restricted our focus to technologies supporting individual mindfulness practices and excluded literature concerning mindfulness communities, collaborative practices, and group mindfulness sessions. As a result, our framework does not currently account for the social and interpersonal dynamics involved in collective mindfulness experiences. Future research should therefore explore how our framework may be expanded to support and account for group settings.

Third, our framework-building process adopted a retrospective approach, drawing on existing work within the research community. Hence, the framework reflects current design practices and conceptualizations of mindfulness technologies. Novel approaches--for example, new methods of embodying the intended object in FAM or innovative techniques for eliciting emotions in OMM--may challenge, extend, or necessitate revisions to our framework. Future iterations should therefore remain open to incorporating emerging design paradigms and novel interaction approaches.

\section{Conclusion}
\label{sec:conclusion}
Throughout this paper, we make four primary contributions. First, we provide a scoping review of existing tangible interactive systems designed to support different mechanisms of mindfulness practice. Second, we review the evaluation methods currently used to assess the effectiveness of tangible interactive systems for mindfulness-related outcomes. Through this review, we identified a lack of theoretical grounding as a key factor limiting the growth and coherence of the field. In response, our third contribution proposes a theoretical framework for concentrative practice in focused-attention meditation (FAM), describing how tangible interactive systems support attentional regulation and body awareness. Fourth, we introduce a theoretical framework for open-monitoring meditation (OMM), outlining how tangible interactive systems support body awareness, emotion regulation, and change in self in perspective. Finally, we discuss key research gaps and propose future directions for the design and evaluation of tangible mindfulness technologies. We hope these contributions provide a more comprehensive theoretical and evaluative foundation for the continued development of tangible mindfulness systems within HCI.




\bibliographystyle{ACM-Reference-Format}
\bibliography{manuscript_refs, scoping_review_ref}
\newpage
\section*{Appendix I - Methodological constraint of Scoping Review}
\label{sec:Appendix_1}

The primary constraint of this review is that our search was restricted to selected HCI-focused publication venues. While this approach allowed us to capture the research most relevant to the HCI community, it may have excluded pertinent work published in adjacent disciplines (e.g., psychology or health) that are not indexed within these venues. In an ideal scenario, a comprehensive search would include broad indexing services such as Google Scholar\footnote{\url{https://scholar.google.com}}, which aggregate literature across disciplines and publication types. However, at the time of our final search (12 May 2026), our query returned over 200,000 results on Google Scholar, rendering exhaustive screening infeasible within the constraints of this study. This challenge is consistent with prior work. For example,~\citet{terzimehicReviewAmpAnalysis2019} limited their review to the ACM Digital Library and employed a point-based prioritization scheme to manage the scale of retrieved records. In contrast, our approach sought to balance breadth and feasibility by focusing on major HCI venues while complementing database searches with iterative inclusion strategies--namely venue expansion and backward/forward citation chaining--following~\citet{arkseyScopingStudiesMethodological2005}. As our primary aim is to map the research contributions most relevant to the HCI community, we consider this trade-off appropriate. Nevertheless, we acknowledge that our findings may not fully capture the entirety of interdisciplinary work on tangible interfaces for mindfulness.

\section*{Appendix III}
\label{sec:Appendix_2}

\begin{table}[H]
\caption{Data Extraction Framework}
\label{tab:data-extraction}
\begin{tabular}{p{0.38\linewidth} p{0.58\linewidth}}
\toprule

\multicolumn{2}{l}{\textbf{General Information}} \\
Study (name) & Name of the study. \\
Type of paper & Publication type (e.g., conference paper, journal article). \\
Author location & Geographic location(s) of the primary author. \\

\addlinespace
\multicolumn{2}{l}{\textbf{Research Objective \& Contribution}} \\
Research question & Primary research question(s) addressed in the study. \\
Main finding & Key results or conclusions reported by the authors. \\
Contribution type & Nature of contribution (e.g., artifact, theoretical, empirical, technological, method, design research, secondary research). \\

\addlinespace
\multicolumn{2}{l}{\textbf{User Evaluation / Empirical Study}} \\
Country & Country where the study or evaluation was conducted. \\
Type of data & Type of data collected (quantitative, qualitative, or none). \\
Setting & Study context (e.g., lab, field, in-the-wild, online). \\
Participants & Number of participants (N = X). \\
Age & Participant age range or mean age, if reported. \\
Gender & Participant gender distribution, if reported. \\
Result & Summary of evaluation results or findings. \\

\addlinespace
\multicolumn{2}{l}{\textbf{Evaluation (Quantitative Only)}} \\
Variables & Independent and dependent variables measured. \\
Outcomes & Quantitative outcomes, metrics, or statistical results. \\

\addlinespace
\multicolumn{2}{l}{\textbf{Human-Centered Information (RQ1)}} \\

Mindfulness mechanism & 
Underlying psychological mechanism targeted e.g., one of  
(A) Attention regulation, 
(B) Body awareness, 
(C) Emotion regulation (non-judgmental reappraisal), 
(D) Emotion regulation (non-reactivity), 
(E) Change in perspective on the self (decentering/detachment). \\

Mindfulness framework & 
Mindfulness framework, model, or definition adopted (e.g., Mindfulness-Based Stress Reduction (MBSR), implicit or self-defined mindfulness). \\

\addlinespace
\multicolumn{2}{l}{\textbf{Tangible Device Information (RQ2)}} \\
Device sensing method & Type of sensing technology used (if any). \\
Device sensing purpose & Purpose or rationale for sensing. \\
Device feedback method & Type of feedback provided (e.g., haptic, visual, auditory). \\
Device feedback purpose & Purpose or rationale for feedback. \\
\bottomrule
\label{appendix:charting_protocol}
\end{tabular}
\end{table}

\section{Appendix IV}
\label{sec:Appendix_3}

Through the focus group, we ask the overarching research question \textit{What design constraints must HCI practitioners consider when engineering those embodied metaphors?}

The focus group was structured into six phases: (1) introduction to the probes (2) identifying challenges faced during MBSR program, (3) mapping the challenges to the teacher \& student perspectives, (4) ideating solutions, (5) selecting viable ideas, and (6) refining selected ideas through discussion. We align these activities with the stages of the double diamond model of design thinking ~\cite{kochanowskaDoubleDiamondModel2021}  to maximize participants' creativity and, importantly, to guide mindfulness experts through the design journey typically undertaken by HCI designers when creating devices for mindfulness.

\subsection{Participants and Ethics}

Participant recruitment was conducted primarily through the university portal and email. For mindfulness experts, we contacted 20 shortlisted mindfulness centers located in proximity to the university, using the email addresses provided on their official websites. The shortlisting process was based on the number of years of experience in MBSR explicitly stated on their respective websites.

A total of 8 participants (6 female, 2 male) were recruited for the focus group (see \autoref{tab:participants}). Participants had diverse backgrounds, including expertise in mindfulness, mindfulness research, HCI research, or recent participation in an MBSR program. Participants were divided into 3 smaller groups. Groups were designed to have one mindfulness expert, one recent MBSR participant and one HCI researcher.  However, due to the absence of one planned MBSR participant, the third group was formed with one mindfulness expert working with an HCI research who had extensive meditation experience.

The in-person study was conducted in a maker space lab, as the environment was considered conducive to fostering creativity ~\cite{drakeThisPlaceGives2003}. At the beginning of the session, participants were given a brief tour of the laboratory, which included demonstrations of 3D printing, soldering, and fabrication processes. Ethical approval for the study was obtained from the University Committee. All participants provided informed written consent for their participation.

\subsection{Study Design}
Participants began the session by reviewing the study information and informing their consent. After an icebreaker activity, we initiated with an introduction to HCI and the double diamond design process. We then proceeded with a series of activities, which are explained below.

\paragraph{Activity 1 - Introduction to the Probes} 
The activity was designed to introduce key concepts identified in our scoping review and to establish mild constraints that kept the discussion focused on tangible devices. First, we presented participants with three popular definitions of mindfulness (see \autoref{tab:definitions}). Then, participants were asked to write their own interpretations of mindfulness on sticky notes. Subsequently, we asked participants to share their interpretations within the group.

To establish a theoretical foundation, we introduced the concepts of Embodied Theory ~\cite{shapiroEmbodiedCognition2019, wilsonSixViewsEmbodied2002} and Cross-Modal Perception Theory ~\cite{slobodenyukCrossmodalAssociationsColor2015}.

Finally, we introduced several recent tangible devices found in the scoping review as design probes (see Table \ref{tab:designProbes}) to stimulate discussion within the group, drawing inspiration from ~\cite{madapuranagarajMindfulnessbasedEmbodiedTangible2024a}.

\begin{table*}[t]
    \centering
    \renewcommand{\arraystretch}{1.3} 
    \setlength{\tabcolsep}{8pt} 
    \begin{tabular}{p{6cm} p{3.5cm} p{4.5cm}} 
        \toprule
        \textbf{Definition} & \textbf{Source} & \textbf{Why We Use It} \\
        \midrule
        Mindfulness is awareness that arises through paying attention, on purpose, non-judgmentally in the present moment, .& Jon Kabat-Zinn ~\cite{kabat_zinnMindfulnessbasedInterventionsContext2003}  
         & Used in therapeutic contexts like MBSR. \\ 
        
        A cognitive process of noticing new things; mindfulness is the opposite of mindlessness.  
         & Langer et al. ~\cite{langerMindingMattersConsequences1989}  
         & Applied in psychology and cognitive research. \\ 
        
        Mindfulness consists of two key components: (1) the ability to regulate attention, focusing on immediate experiences; and (2) adopting a mindset of curiosity, openness, and acceptance toward present experiences.  
        & Bishop et al. ~\cite{bishopMindfulnessProposedOperational2004}  
        & A testable operational definition of mindfulness \\ 
        \bottomrule
    \end{tabular}
    \caption{Definitions of Mindfulness presented in the focus group}
    \label{tab:definitions}
\end{table*}

        


    

\begin{table*}[t]
    \centering
    \begin{tabular}{l l}
        \toprule
        \textbf{Source} & \textbf{Image} \\
        \midrule
        Manifesting Breath~\cite{farrallManifestingBreathEmpirical2023a}
            & \textit{Abstract:}\ \includegraphics[scale=0.3]{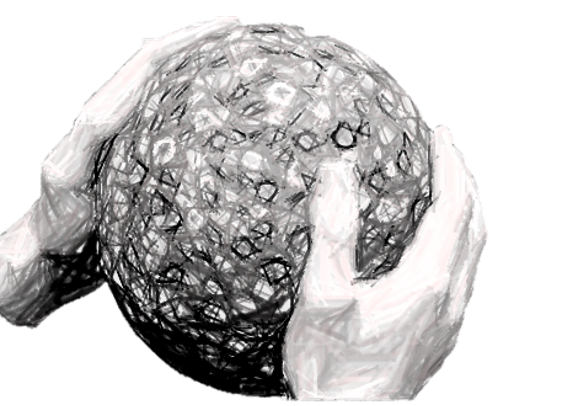} \\[0.5em]
            
        Deep Touch~\cite{jungExploringAwarenessBreathing2021a}
            & \textit{Abstract:}\ \includegraphics[scale=0.4]{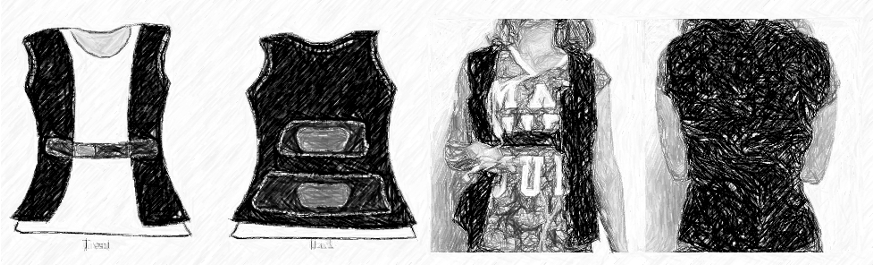} \\[0.5em]
        
        Interoceptive Interaction\cite{daudenroquetInteroceptiveInteractionEmbodied2021a}
            & \textit{Abstract:}\ \includegraphics[scale=0.4]{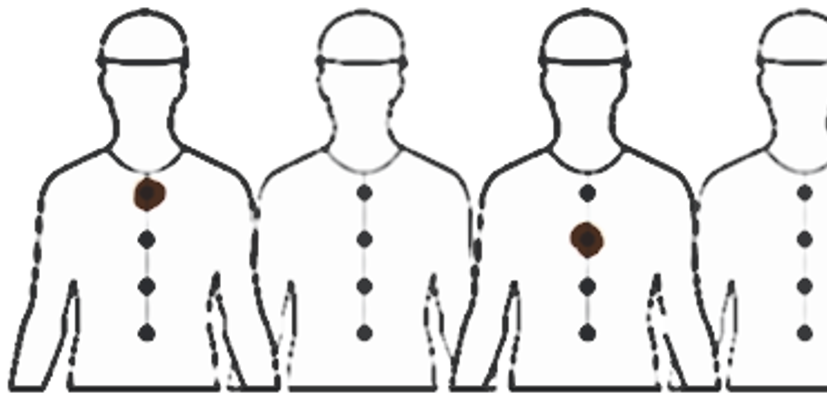} \\
        \bottomrule
    \end{tabular}
    \caption{HCI literature on mindfulness used as design probes.}
    \label{tab:designProbes}
\end{table*}

\paragraph{Activity 2 - Identifying challenges: Who, What, Where, Why} 
This activity was designed to identify the potential challenges faced by mindfulness practitioners. Taking the perspective of \emph{mindfulness teachers} (instructing students) or \emph{mindfulness students} (learning mindfulness), participants were asked to answer the following questions on post-its:
\begin{itemize}
    \item What are your key challenges when teaching/learning mindfulness/meditation?
    \item How do you currently address resistance or disengagement?
    \item What was the most challenging part of practicing mindfulness?
    \item Did you experience challenges in maintaining a regular practice?
\end{itemize}

The resulting post-its with challenges were used in the next stage to map out the problem space.

\paragraph{Activity 3 - Mapping the challenges and solutions: Mind Map} 
In this activity, participants generated solutions to the problem space identified in the previous task. They were then instructed to construct a mind map using the \textit{''teacher''} and\textit{ ''student''} as the anchor points, situating the solution space in relation to the identified challenges and their corresponding solutions.

Participants then discussed their challenges and solutions across groups before moving into ideation (Activity 4). This group exchange encouraged them to brainstorm mindfulness devices based on both individual and shared ideas.

\paragraph{Activity 4 - Ideation and prioritizing} 
Participants were introduced to a rapid ideation technique~\cite{banfieldDesignSprintPractical2015} where each individual brainstormed eight ideas for devices that explore how tangible interactions could enhance mindfulness practices within an eight-minute timeframe (one idea per minute).

Participants were asked to create simple sketches (or annotations) to capture their ideas on post-it notes. They were also given idea cards to scaffold their ideas and encourage new ones. 

Each card was designed to depict a single haptic interaction found from the scoping review, including touch, vibration, pressure, temperature, and balance, along with a brief description and example. When designing a tool, participants were instructed to use only one card per idea (e.g., focusing on touch-based vibration, but not combining touch and temperature) to keep the design simple. They were also encouraged to prioritize a number of ideas (breadth) over quality of an idea (depth) to maintain momentum.

After brainstorming, participants shared their eight ideas within the group. To prioritize a single design, we employed a coin-based investment strategy to ensure that no participant's efforts were overlooked in the prioritization process ~\cite{royCardbasedDesignTools2019}. Each participant received \$10 worth of PipeCoin (a virtual currency) and was asked to invest in an idea, treating it as if it were a company or product available for purchase. The prototype idea with the highest investment was selected for the next activity.

\paragraph{Activity 5 - Refining the best idea through discussion} 

In the final activity, the participants thoroughly discussed the prototype idea with the highest investment. To scaffold the conversation, they were provided with a set of discussion points that focused on the prototype selection, its usability, and how these aspects align with definitions of mindfulness and embodied theory. Additionally, the discussion points addressed the challenges HCI researchers face when developing prototypes, such as whether interaction with the device induces performance anxiety and how to foster a sense of non-judgment during interactions, which were identified as critical aspects of mindfulness-related devices in our scoping review.

\subsection{Analysis}



The audio recordings were transcribed with Otter\footnote{\url{https://otter.ai}}, where the first author verified their accuracy by listening to the recordings. Personally identifiable information, including names, was then anonymized and replaced with pseudonyms. Pseudonyms followed a <group ID participant ID> convention (e.g., G1M1). The transcripts and participant drawings were imported into a shared Miro\footnote{\url{https://miro.com}} board for collaborative analysis.

Two coders independently familiarized themselves with the data and generated tentative inductive codes~\cite{terryEssentialsThematicAnalysis2021}. In the first meeting, they reviewed the transcripts from Group 1 line by line, comparing their initial codes and co-constructing a shared code set. The unit of analysis was a complete participant utterance to preserve contextual integrity. In the second meeting, they jointly reviewed and coded the transcripts and sketches from Group 2, refining the code set.
In the third meeting, they consolidated tentative themes into a comprehensive codebook. All versions of the evolving code sets were systematically documented as tables within the Miro board to maintain a clear versioning history. 
Following the completion of coding, the coders reconvened to generate candidate themes, refine themes individually, and subsequently finalize them through discussion.
To ensure accuracy, all selected quotes were re-checked against the original recordings by the first author before deleting the audio raw data.

\paragraph{Coder Positionality}

The first and second authors were involved in thematic analysis. Both coders have a background in Theravada Buddhist meditation practices, and the second coder has further experience in yogic meditation. While these experiences help foster a reflexive stance, they may also incline coders to associate 'mindfulness' and 'meditation' with a positive connotation and value these concepts positively. Additionally, both coders were aware that the first coder's training as an engineer may create a tendency to favor technological-solution-oriented perspectives, such as design, evaluation, and feasibility of meditation tools, whilst, second coder's experience as a designer with qualitative user-centered research expertise may ground their attentiveness to understand user perspectives and lived experiences of the participants rather than looking data with a solution oriented perspective. Therefore, to alleviate the potential biases, both authors engaged in ongoing reflexive practices, including peer discussions, critically examining the assumptions and the rationale for interpretations, and remaining open to participants' diverse experiences.

\end{document}